\documentclass[journal]{IEEEtran}

\pdfoutput=1

\ifCLASSINFOpdf
   \usepackage[pdftex]{graphicx}
\else
\fi
\usepackage{algorithmic}
\usepackage{algorithm}

\usepackage{amsmath,mathtools}
\usepackage{amssymb}
\usepackage{amsfonts}
\usepackage{graphicx}
\usepackage{amsbsy}
\usepackage[margin = 10pt]{caption}
\usepackage[caption=false,font=footnotesize]{subfig}

\usepackage{pdflscape}
\usepackage{rotating}
\usepackage{setspace}
\usepackage{enumitem}
\usepackage{pgfgantt}
\usetikzlibrary{arrows,positioning,shapes,calc,fit}
\usepackage{afterpage}
\usepackage{bm}

\newcommand{\Target}{p}

\newcommand{\NbPart}{N}

\newcommand{\Burnin}{N_b}

\newcommand{\state}{{\bm  x}}
\newcommand{\stateR}{{\state}}
\newcommand {\vc} [1] {\boldsymbol{#1}}
\begin{document}
%
\title{Sequential Markov Chain Monte Carlo for Bayesian Filtering with Massive Data}

\author{Allan~De~Freitas$^{1,{*}}$, Fran\c{c}ois Septier$^2$, Lyudmila~Mihaylova$^1$
\thanks{$^1$ Department of Automatic Control and Systems Engineering, University of Sheffield, United Kingdom.}
\thanks{$^2$ Institute Mines Telecom/Telecom Lille, CRIStAL UMR CNRS 9189, France}
\thanks{$^* Corresponding \, author $}
}
\maketitle

\begin{abstract}
Advances in digital sensors, digital data storage and communications have resulted in systems being capable of accumulating large collections of data. In the light of dealing with the challenges that massive data present, this work proposes solutions to inference and filtering problems within the Bayesian framework. Two novel Bayesian inference algorithms are developed for non-linear and non-Gaussian state space models, able to deal with large volumes of data (or observations). These are sequential Markov chain Monte Carlo (MCMC) approaches relying on two key ideas: 1) subsample the massive data and utilise a smaller subset for filtering and inference, and  2) a divide and conquer type approach computing local filtering distributions each using a subset of the measurements. Simulation results highlight the accuracy and the large computational savings, that can reach 90\% by the proposed algorithms when compared with standard techniques.
\end{abstract}


%
\IEEEpeerreviewmaketitle

\section{Introduction}

In many applications, it is of interest to estimate a signal from a sequence of collected data. In a Bayesian framework, this involves the sequential inference of the filtering distribution associated with a state space model. The solution is referred to as the Kalman filter \cite{Kalman1960} when the state space model is linear and Gaussian. However, there is typically no analytically tractable solution when the state space model is non-linear and/or non-Gaussian. Several algorithms which achieve sequential inference in such systems through approximations have been proposed.

One such class of techniques are referred to as sequential Monte
Carlo (SMC) methods \cite{Cappe2007}, or particle filters (PFs), which
involves a weighted discrete approximation of the filtering
distribution, and utilise importance sampling. PFs have been
successfully applied to many areas.

Research on efficient implementations of SMC methods have focused on making the structure of the PF parallel \cite{Bolic2005}, particularly the resampling step \cite{Li2015}, which can then be used in distributed processing applications \cite{Read2014}. However, this typically requires approximations to achieve a solution and still requires the evaluation of all the data. Also in very high dimensional problems and massive data, the PF is prone to weight degeneracy and sample impoverishment \cite{Bengtsson2008,VanLeeuwen2014}.

A related but promising alternative to PFs is the sequential Markov chain Monte Carlo (MCMC) method \cite{Khan2005,Septier2009}, which has been successfully applied in several challenging areas \cite{Lamberti2015}. The sequential MCMC method does not rely on importance sampling and instead utilises the power of MCMC techniques in a sequential setting to perform inference. Analysing massive amounts of data with sequential MCMC can lead to long processing times which are problematic in time sensitive filtering applications.

In static MCMC simulation, there have been several different approaches proposed for dealing with large amounts of data \cite{Bardenet2015a}. The proposed methods can be categorised as either parallel or iterative strategies.

In terms of parallel strategies, there are two general approaches which have been proposed. The first approach is referred to as blocking. These techniques focus on parallelising specific steps in the MCMC approach. In \cite{Suchard2010} it was proposed to parallelise the computation of the likelihood. This is restrictive in terms of the model used, and requires a large amount of communication between the processors. The second approach is referred to as divide and conquer. Techniques based on divide and conquer focus on subdividing the measurements and running separate MCMC samplers in parallel on each subdivided set of measurements. The samples from the separate MCMC samplers, referred to as local samples, are then combined to obtain samples from the complete posterior distribution, referred to as global samples. The divide and conquer techniques differ in how the local samples are combined to obtain the global samples. In \cite{Scott2013}, global samples are obtained as a weighted average of the local samples. This approach is only theoretically valid under a Gaussian assumption. In \cite{Neiswanger2013}, the local posterior from the separate MCMC samplers is approximated as Gaussian or with a Gaussian kernel density estimation. Global samples can then be obtained through the product of the local densities. This work was further extended for time series analysis in \cite{Casarin2015}. This idea is also further developed in \cite{Wang2013} by representing the discrete kernel density estimation as a continuous Weierstrass transform. In \cite{Minsker2014}, the combination is based on the geometric median of the local posteriors which are approximated with Weiszfeld's algorithm by embedding the local posteriors in a reproducing kernel Hilbert space. Divide and conquer techniques typically struggle in applications where the local posteriors substantially differ, and if they do not satisfy Gaussian assumptions. In \cite{Xu2014,Gelman2014} a divide and conquer strategy was proposed which attempts to overcome the challenge of differing local posteriors, and relaxing the Gaussian assumption to a more general assumption of a posterior distribution from the exponential family. The approach is based on the expectation propagation algorithm. In this approach, the separate MCMC samplers exchange sufficient statistics, resulting in each individual MCMC sampler converging to the global posterior.

Iterative strategies rely on subsampling mechanisms, such as pseudo likelihoods \cite{Andrieu2009,Quiroz2014} or confidence intervals \cite{Bardenet2014,Korattikara2014}, to perform inference using MCMC techniques based only on a certain subsample of all the measurements.

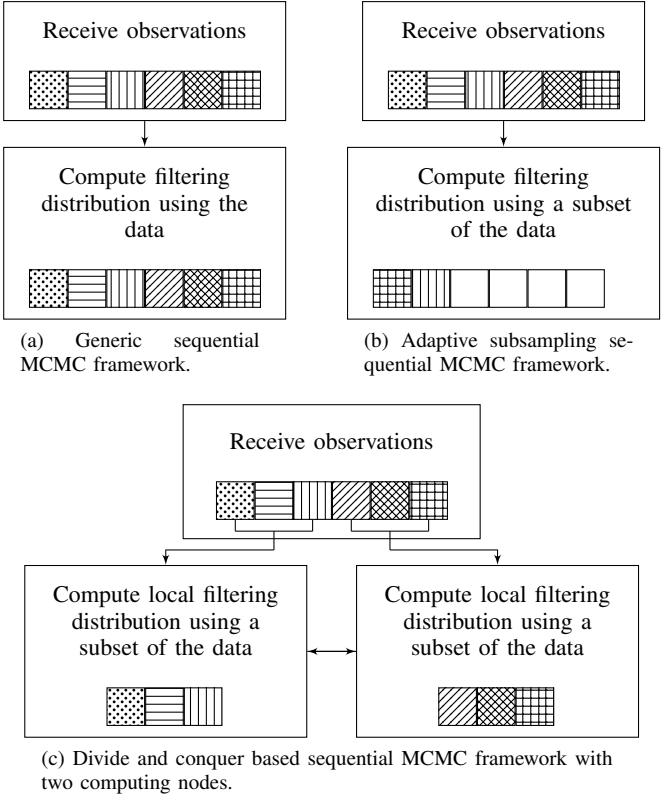
\begin{figure}
\centering
\subfloat[Generic sequential MCMC framework.]{
\begin{tikzpicture}[node distance=10mm, >=latex',
 block/.style = {draw, rectangle, minimum height=5mm, minimum width=5mm,align=center},
                        ]
    \node  [] (MTT) {\small \begin{minipage}{32mm}\center Receive observations \end{minipage}};
    \node [block, below left= 3mm and -7mm of MTT,pattern=crosshatch dots] (M1)  {};
     \node [block, right= 0mm of M1,pattern=horizontal lines] (M2)  {};
      \node [block, right= 0mm of M2,pattern=vertical lines] (M3)  {};
       \node [block, right= 0mm of M3,pattern=north east lines] (M4)  {};
        \node [block, right= 0mm of M4,pattern=crosshatch] (M5)  {};
         \node [block, right= 0mm of M5,pattern=grid] (M6)  {};
    \node [fit= (MTT) (M1) (M2) (M3) (M4) (M5) (M6), draw, inner sep=0.15cm] (BGT)  {};
    
    \node  [below=5mm of BGT] (MTT2) {\small \begin{minipage}{32mm}\center Compute filtering distribution using the data \end{minipage}};
    \node [block, below left= 3mm and -7mm of MTT2,pattern=crosshatch dots] (M12)  {};
     \node [block, right= 0mm of M12,pattern=horizontal lines] (M22)  {};
      \node [block, right= 0mm of M22,pattern=vertical lines] (M32)  {};
       \node [block, right= 0mm of M32,pattern=north east lines] (M42)  {};
        \node [block, right= 0mm of M42,pattern=crosshatch] (M52)  {};
         \node [block, right= 0mm of M52,pattern=grid] (M62)  {};
    \node [fit= (MTT2) (M12) (M22) (M32) (M42) (M52) (M62), draw, inner sep=0.15cm] (BGT2)  {};
    \path[draw,->] (BGT.south) -| (BGT2.north);       
    \end{tikzpicture}}
    \hspace*{\fill}%
\subfloat[Adaptive subsampling sequential MCMC framework.]{
\begin{tikzpicture}[node distance=10mm, >=latex',
 block/.style = {draw, rectangle, minimum height=5mm, minimum width=5mm,align=center},
                        ]
    \node  [] (MTT) {\small \begin{minipage}{32mm}\center Receive observations \end{minipage}};
    \node [block, below left= 3mm and -7mm of MTT,pattern=crosshatch dots] (M1)  {};
     \node [block, right= 0mm of M1,pattern=horizontal lines] (M2)  {};
      \node [block, right= 0mm of M2,pattern=vertical lines] (M3)  {};
       \node [block, right= 0mm of M3,pattern=north east lines] (M4)  {};
        \node [block, right= 0mm of M4,pattern=crosshatch] (M5)  {};
         \node [block, right= 0mm of M5,pattern=grid] (M6)  {};
    \node [fit= (MTT) (M1) (M2) (M3) (M4) (M5) (M6), draw, inner sep=0.15cm] (BGT)  {};
    
    \node  [below=5mm of BGT] (MTT2) {\small \begin{minipage}{36mm}\center Compute filtering distribution using a subset of the data \end{minipage}};
    \node [block, below left= 3mm and -7mm of MTT2,pattern=grid] (M12)  {};
     \node [block, right= 0mm of M12,pattern=vertical lines] (M22)  {};
      \node [block, right= 0mm of M22] (M32)  {};
       \node [block, right= 0mm of M32] (M42)  {};
        \node [block, right= 0mm of M42] (M52)  {};
         \node [block, right= 0mm of M52] (M62)  {};
    \node [fit= (MTT2) (M12) (M22) (M32) (M42) (M52) (M62), draw, inner sep=0.15cm] (BGT2)  {};
    \path[draw,->] (BGT.south) -|  (BGT2.north);       
    \end{tikzpicture}}
\qquad
\vspace*{\fill}%
\subfloat[Divide and conquer based sequential MCMC framework with two computing nodes.]{
\begin{tikzpicture}[node distance=10mm, >=latex',
 block/.style = {draw, rectangle, minimum height=5mm, minimum width=5mm,align=center},
                        ]
    \node  [] (MTT) {\small \begin{minipage}{32mm}\center Receive observations \end{minipage}};
    \node [block, below left= 3mm and -7mm of MTT,pattern=crosshatch dots] (M1)  {};
     \node [block, right= 0mm of M1,pattern=horizontal lines] (M2)  {};
      \node [block, right= 0mm of M2,pattern=vertical lines] (M3)  {};
       \node [block, right= 0mm of M3,pattern=north east lines] (M4)  {};
        \node [block, right= 0mm of M4,pattern=crosshatch] (M5)  {};
         \node [block, right= 0mm of M5,pattern=grid] (M6)  {};
    \node [fit= (MTT) (M1) (M2) (M3) (M4) (M5) (M6), draw, inner sep=0.25cm] (BGT)  {};
     
    \node  [below right= 5mm and -15mm of BGT] (MTT21) {\small \begin{minipage}{32mm}\center Compute local filtering distribution using a subset of the data \end{minipage}};
    \node [block, below left=3mm and -14.5mm of MTT21,pattern=north east lines] (M121)  {};
     \node [block, right= 0mm of M121,pattern=crosshatch] (M221)  {};
      \node [block, right= 0mm of M221,pattern=grid] (M231)  {};
    \node [fit= (MTT21) (M121) (M221) (M231), draw, inner sep=0.15cm] (BGT21)  {};
    
    \node  [below left= 5mm and -15mm of BGT] (MTT22) {\small \begin{minipage}{32mm}\center Compute local filtering distribution using a subset of the data \end{minipage}};
    \node [block,below left=3mm and -14.5mm of MTT22,pattern=crosshatch dots] (M122)  {};
     \node [block, right= 0mm of M122,pattern=horizontal lines] (M222)  {};
      \node [block, right= 0mm of M222,pattern=vertical lines] (M322)  {};
    \node [fit= (MTT22) (M122) (M222) (M322), draw, inner sep=0.15cm] (BGT22)  {};
    
        \coordinate (A) at ($ (M1.south) + (0,-1.5mm) $);
    \coordinate (B) at ($ (M3.south) + (0,-1.5mm) $);
    \coordinate (C) at ($ (M2.south) + (0,-1.5mm) $);
    \coordinate (D) at ($ (M2.south) + (0,-4mm) $);
     \coordinate (E) at ($ (BGT22.north) + (0,2mm) $);
    \path[draw] (M1.south) -| (A);
     \path[draw] (M3.south) -| (B);
     \path[draw] (A) -| (B);
     \path[draw] (C) -| (D);
        \path[draw] (D) -| (E);
        \path[draw,->] (E) -| (BGT22.north);
         \coordinate (A1) at ($ (M4.south) + (0,-1.5mm) $);
    \coordinate (B1) at ($ (M6.south) + (0,-1.5mm) $);
    \coordinate (C1) at ($ (M5.south) + (0,-1.5mm) $);
    \coordinate (D1) at ($ (M5.south) + (0,-4mm) $);
     \coordinate (E1) at ($ (BGT21.north) + (0,2mm) $);
    \path[draw] (M4.south) -| (A1);
     \path[draw] (M6.south) -| (B1);
     \path[draw] (A1) -| (B1);
      \path[draw] (C1) -| (D1);
        \path[draw] (D1) -| (E1);
        \path[draw,->] (E1) -| (BGT21.north);
    
    \path[draw,<-] (BGT22.east) -| (BGT21.west);
    \path[draw,<-] (BGT21.west) -| (BGT22.east);
    \end{tikzpicture}}
\caption{Comparison of the proposed sequential MCMC frameworks to the generic sequential MCMC framework at each discrete time step.}\label{frameworkcompare}
\end{figure}

The key contributions of this work are in the proposed solutions for inference and filtering problems with the Bayesian framework. Two novel sequential MCMC algorithms for dealing with massive data are introduced. The algorithms achieve computational efficiency while maintaining accurate estimates. The first algorithm achieves this through the introduction of adaptive subsampling in the sequential MCMC framework, preliminary results of which are introduced in \cite{DeFreitas2015}; and the second algorithm by merging the expectation propagation and sequential MCMC frameworks. The structure of the algorithms is compared in Figure \ref{frameworkcompare}. The performance of the algorithms is explored through two detailed examples.

\section{Problem Formulation}
The primary distribution of interest in a Bayesian framework is the filtering distribution $p(\vc{x}_k|\vc{z}_{1:k})$, where $\vc{x}_k \in \mathbb{R}^{n_{\vc{x}}}$ is the state vector at time $t_k$ with $k = {1,...,T} \in \mathbb{N}$, and $\vc{z}_{1:k} = \{\vc{z}_1,...,\vc{z}_k\}$, represents all the data received up till time $t_k$. The data received at each time $t_k$ are represented by a set $\vc{z}_k = \{\vc{z}_k^1,...,\vc{z}_k^{M_k}\}$, where $M_k$ is the total amount of data and $\vc{z}_k^i \in \mathbb{R}^{n_{\vc{z}}}$. In this paper the data is considered independent. The filtering distribution can be recursively updated based on
\begin{equation}
p(\vc{x}_k|\vc{z}_{1:k}) \propto \int p(\vc{z}_k|\vc{x}_k) p(\vc{x}_k|\vc{x}_{k-1})p(\vc{x}_{k-1}|\vc{z}_{1:k-1})d\vc{x}_{k-1},\label{filterpdf}
\end{equation}
where $p(\vc{z}_k|\vc{x}_k)$ is referred to as the likelihood probability density function (pdf), and $p(\vc{x}_k|\vc{x}_{k-1})$ is referred to as the state transition pdf. An analytical solution to \eqref{filterpdf} is typically intractable when the state space model is characterised by non-linearities and/or non-Gaussian noise.
\subsection{Sequential Markov Chain Monte Carlo}
MCMC methods work by constructing a Markov chain with a desired distribution as the equilibrium distribution. A common MCMC technique used to obtain samples from the equilibrium distribution, $\pi(\vc{x})$, is the Metropolis-Hastings (MH) algorithm. This is achieved by first generating a sample from a known proposal distribution $\vc{x}^*\sim q(\,\cdot\,|\vc{x}^{m-1})$. The proposed sample is accepted as the current state of the chain, $\vc{x}^m$, with probability
\begin{equation}
\rho = \min\left(1, \frac{\pi(\vc{x}^*)q(\vc{x}^{m-1}|\vc{x}^*)}{\pi(\vc{x}^{m-1})q(\vc{x}^*|\vc{x}^{m-1})}\right),\label{eq:acceptance_prob}
\end{equation}
where $\rho$ is referred to as the acceptance probability, otherwise the previous state of the chain is stored as the current state, $\vc{x}^m = \vc{x}^{m-1}$.

In \cite{Khan2005} it was proposed to use MCMC methods, specifically the MH algorithm, to target the filtering distribution in \eqref{filterpdf} as the equilibrium distribution. This allows for the iterative update of an approximation of the filtering distribution by representing $p(\vc{x}_{k-1}|\vc{z}_{1:k-1})$ with a set of unweighted particles,
\begin{equation}
p(\vc{x}_{k-1}|\vc{z}_{1:k-1}) \approx \frac{1}{N}\sum_{j=1}^{N}\delta(\vc{x}_{k-1}-\vc{x}_{k-1}^{(j)}),
\end{equation}
 where $N$ is the number of particles, $\delta(\,\cdot\,)$ denotes the Dirac delta function, and $(j)$ the particle index. This technique was shown to work well in state space models containing a high number of dimensions when compared to techniques relying on importance sampling, however, this direct approach may result in a high computational expense \cite{Septier2009}.

It was proposed in \cite{Septier2009} to consider targeting the joint filtering distribution of $\vc{x}_k$ and $\vc{x}_{k-1}$,
\begin{equation}
p(\vc{x}_k,\vc{x}_{k-1}|\vc{z}_{1:k}) \propto  p(\vc{z}_k|\vc{x}_k) p(\vc{x}_k|\vc{x}_{k-1})p(\vc{x}_{k-1}|\vc{z}_{1:k-1}),\label{eq:jointposterior}
 \end{equation}
as the equilibrium distribution in order to help alleviate the high computational demand. In a similar fashion, an approximation for the joint filtering distribution can be obtained through MCMC methods by representing $p(\vc{x}_{k-1}|\vc{z}_{1:k-1})$ with a set of unweighted particles. This approach has the advantage of avoiding the direct Monte Carlo computation of the predictive posterior density. Furthermore, the approximation can be marginalised to obtain the filtering distribution of interest. It was also proposed in \cite{Septier2009}, to utilise a composite MCMC kernel to generate a Markov chain with stationary distribution described by \eqref{eq:jointposterior}. The composite MCMC kernel is based on both joint and conditional draws and has been shown to be more efficient in high dimensional systems \cite{Septier2009,Mihaylova2014}.

More specifically, the composite MCMC kernel is comprised of two steps. The first step is comprised of a joint draw for $\vc{x}_k$ and $\vc{x}_{k-1}$ through the application of a MH sampler with proposal distribution $q_1(\,\cdot\,)$. The second step, referred to as the refinement step, draws $\vc{x}_{k-1}$ given the current state of the Markov chain for $\vc{x}_k$ with proposal distribution $q_2(\,\cdot\,)$, followed by a draw of $\vc{x}_k$ given the current state of the Markov chain for $\vc{x}_{k-1}$ with proposal distribution $q_3(\,\cdot\,)$. Furthermore, when $\vc{x}_k$ is high dimensional, a series of block MH within Gibbs steps can be used to update it efficiently. The refinement step is introduced to aid in the mixing of the chain. An appropriate burn in period, $N_{b}$, was also introduced to minimize the effect of the initial values of the Markov chain. The implementation of this procedure is referred to as the generic sequential MCMC algorithm and is summarised in Algorithm \ref{alg1}. When considering massive amounts of data, the computation of the likelihood becomes excessively expensive, rendering this approach infeasible.
 \begin{algorithm}[!ht]
\caption{Generic Sequential Markov Chain Monte Carlo}
\label{alg1}
\begin{algorithmic}[1]
\small
\STATE Initialize particle set: $\{\vc{x}_0^{(j)}\}_{j=1}^{N}$
\FOR{$k$ = 1,...,$T$}
\FOR{$m$ = 1,...,$N+N_b$}
\STATE \textit{\underline{Joint Draw}}
\STATE Propose $\{\vc{x}_k^*,\vc{x}_{k-1}^*\} \sim q_1\left(\vc{x}_k,\vc{x}_{k-1}|\vc{x}_k^{m-1},\vc{x}_{k-1}^{m-1}\right)$
\STATE Compute the MH acceptance probability $\rho_1 =$ \\ $\min \left(1,\frac{p(\vc{x}^*_k,\vc{x}^*_{k-1}|\vc{z}_{1:k}) }{q_1\left(\vc{x}^*_k,\vc{x}^*_{k-1}|\vc{x}_k^{m-1},\vc{x}_{k-1}^{m-1}\right)}\frac{q_1\left(\vc{x}_k^{m-1},\vc{x}_{k-1}^{m-1}|\vc{x}^*_k,\vc{x}^*_{k-1}\right) }{p(\vc{x}^{m-1}_k,\vc{x}^{m-1}_{k-1}|\vc{z}_{1:k})} \right)$
\STATE Accept $\{\vc{x}_k^m,\vc{x}_{k-1}^m\} = \{\vc{x}_k^*,\vc{x}_{k-1}^*\}$ with probability $\rho_1$
\STATE \textit{\underline{Refinement}}
\STATE Propose $\{\vc{x}_{k-1}^*\} \sim q_2\left(\vc{x}_{k-1}|\vc{x}_k^{m},\vc{x}_{k-1}^{m}\right)$
\STATE Compute the MH acceptance probability $\rho_2 =$ \\ $\min \left(1,\frac{p(\vc{x}^*_{k-1}|\vc{x}^m_k,\vc{z}_{1:k}) }{q_2\left(\vc{x}^*_{k-1}|\vc{x}_k^{m},\vc{x}_{k-1}^{m}\right)}\frac{q_2\left(\vc{x}_{k-1}^{m}|\vc{x}^m_k,\vc{x}^*_{k-1}\right) }{p(\vc{x}^{m}_{k}|\vc{x}^{m}_{k-1},\vc{z}_{1:k})} \right)$
\STATE Accept $\{\vc{x}_{k-1}^m\} = \{\vc{x}_{k-1}^*\}$ with probability $\rho_2$
\STATE Divide $\vc{x}_k$ into $P$ disjoint blocks $\{\Omega_p\}_{p=1}^P$ such that \\ $\bigcup_p \Omega_p=\{1,...,N_d\}$ and $\Omega_p \cap \Omega_q = \emptyset, \forall p \neq q$
\FOR{$p$ = $1$,...,$P$}
\STATE Propose $\{\vc{x}_k^*(\Omega_p)\} \sim q_{3,p}\left(\vc{x}_k(\Omega_p)|\vc{x}_k^{m},\vc{x}_{k-1}^{m}\right)$
\STATE Compute the MH acceptance probability $\rho_{3,p} =$ \\ $\min \left(1,\frac{p(\vc{x}^*_{k}(\Omega_p)|\vc{x}^m_{k-1},\vc{z}_{1:k}) }{q_{3,p}\left(\vc{x}^*_{k}(\Omega_p)|\vc{x}_k^{m},\vc{x}_{k-1}^{m}\right)}\frac{q_{3,p}\left(\vc{x}_{k}^{m}(\Omega_p)|\vc{x}^*_k,\vc{x}^m_{k-1}\right) }{p(\vc{x}^{m}_{k}(\Omega_p)|\vc{x}^{m}_{k-1},\vc{z}_{1:k})} \right)$
\STATE Accept $\{\vc{x}_{k}^m(\Omega_p)\} = \{\vc{x}_{k}^*(\Omega_p)\}$ with probability $\rho_{3,p}$
\ENDFOR
\ENDFOR
\STATE Approximation of the marginal posterior distribution with the following empirical measure: \\ $\hat{p}(\vc{x}_k|\vc{z}_{1:k}) = \frac{1}{N}\sum_{j=N_b + 1}^{N+N_b}\delta(\vc{x}_k-\vc{x}_k^{(j)})$
\ENDFOR
\normalsize
\end{algorithmic}
\end{algorithm}
\section{Adaptive Subsampling Sequential Markov chain Monte Carlo}
In the generic sequential MCMC algorithm, calculating the acceptance probabilities, $\rho_1$ and $\rho_{3,p}$, requires the evaluation of all the measurements. In this section we merge the concept of adaptive subsampling to sequential MCMC to reduce this computational burden.

Looking back at the standard MH sampler, equation \eqref{eq:acceptance_prob} can be interpreted as the acceptance of the  proposed sample, $\vc{x}^*$, as the current state of the chain, $\vc{x}^m$, if the following condition is satisfied
\begin{equation}
u < \frac{\pi(\vc{x}^*)q(\vc{x}^{m-1}|\vc{x}^*)}{\pi(\vc{x}^{m-1})q(\vc{x}^*|\vc{x}^{m-1})},\end{equation}
where $u$ represents a sample from a uniform distribution $u \sim \textit{U}_{[0,1]}$. This expression can be further developed by applying Bayes' rule and assuming that there are $M$ conditionally independent measurements, $\vc{z}^i$:
\begin{equation}
u < \frac{p(\vc{x}^*)q(\vc{x}^{m-1}|\vc{x}^*)}{p(\vc{x}^{m-1})q(\vc{x}^*|\vc{x}^{m-1})}\prod_{i=1}^M \frac{p(\vc{z}^i|\vc{x}^*)}{p(\vc{z}^i|\vc{x}^{m-1})}.
\end{equation}
The previous state of the chain is stored as the current state, $\vc{x}^m = \vc{x}^{m-1}$, when the proposed sample does not meet this criterion. We further manipulate this expression into a form with the likelihoods isolated:\small
\begin{align}
\frac{1}{M}\log\left[u \frac{p(\vc{x}^{m-1})q(\vc{x}^*|\vc{x}^{m-1})}{p(\vc{x}^*)q(\vc{x}^{m-1}|\vc{x}^*)}\right] &<\frac{1}{M}\sum_{i=1}^M \log\left[\frac{p(\vc{z}^i|\vc{x}^*)}{p(\vc{z}^i|\vc{x}^{m-1})}\right], \nonumber \\ \psi(\vc{x}^{m-1},\vc{x}^*) &<\Lambda^M(\vc{x}^{m-1},\vc{x}^*) .
\end{align}
\normalsize
When the number of measurements is very large, the log likelihood ratio becomes the most computationally expensive part of the generic sequential MCMC algorithm. To reduce the computational complexity, a Monte Carlo (MC) approximation for the log likelihood ratio has been proposed \cite{Bardenet2015}:
\begin{equation}
\Lambda^{S_{m}}(\vc{x}^{m-1},\vc{x}^*) = \frac{1}{S_m}\sum_{i=1}^{S_m} \log\left[\frac{p(\vc{z}^i|\vc{x}^*)}{p(\vc{z}^i|\vc{x}^{m-1})}\right]\label{MCappr}
\end{equation}
where the set $\vc{z}^* = \{\vc{z}^{1,*},...,\vc{z}^{S_{m},*}\}$ is drawn uniformly without replacement from the original set of $M$ measurements.

The difficulty which arises is in selecting a minimum value for $S_{m}$ that results in a set of subsampled measurements that contain enough information to make the correct decision in the MH step. To overcome this difficulty in standard MCMC for static inference, the authors in \cite{Bardenet2014} proposed to use concentration inequalities which provide a probabilistic bound on how functions of independent random variables deviate from their expectation. In this case, the independent random variables are the log likelihood ratio terms. Thus, it is possible to obtain a bound on the deviation of the MC approximation in \eqref{MCappr} from the complete log likelihood ratio:
\begin{equation}
P(|\Lambda^{S_{m}}_1(\vc{x}^{m-1},\vc{x}_k^*)-\Lambda^{M}_1(\vc{x}^{m-1},\vc{x}^*)|\! \leq c_{S_{m}})\! \geq \! 1-\delta_{S_{m}}
\end{equation}
where $\delta_{S_{m}} >0$, and $c_{S_{m}}$ is dependent on which inequality is used. There are several inequalities which could be used, in this paper we make use of the  empirical Bernstein inequality \cite{audibert2009, Bardenet2015}, which results in:
\begin{equation}
c_{S_{m}} = \sqrt{\frac{2V_{S_{m}}\log({3/ \delta_{S_{m}}})}{{S_{m}}}} + \frac{3R\log({3/ \delta_{S_{m}}})}{{S_{m}}} \label{ct}
\end{equation}
where $V_{S_{m}}$ represents the sample variance of the log likelihood ratio, and $R$ is the range given by
\begin{eqnarray}
\nonumber R=\max_{1\le i \le {M}}\!\left\{\!\log\!\left[ \frac{p(\vc{z}^{i}|\vc{x}^*)}{p(\vc{z}^{i}|\vc{x}^{m-1})}\right]\!\right\}-\\\min_{1\le i \le {M}}\!\left\{\log\! \left[\frac{p(\vc{z}^{i}|\vc{x}^*)}{p(\vc{z}^{i}|\vc{x}^{m-1})}\right]\right\}\!\!\!\!\!\!\!\!\!\!\!\!\!\!\!\!\!\!\!\!\!\!\!\!
\end{eqnarray}

Looking back at the standard sequential MCMC approach, we find that the joint draw is accepted based on the condition $\Lambda^{M}(\vc{x}^*,\vc{x}^{m-1})>\psi(\vc{x}^*,\vc{x}^{m-1})$. It is required to relate this expression in terms of the MC approximation of \eqref{MCappr}. Since the MC approximation is bounded, we can state that it is not possible to make a decision when the value of $\psi(\vc{x}^*\vc{x}^{m-1})$ falls within the region specified by the bound. Thus it is required that $|\Lambda^{S_{m}}(\vc{x}^{m-1},\vc{x}^*)-\psi(\vc{x}^*,\vc{x}^{m-1})|>c_{S_{m}}$, where $|\,\cdot\,|$ represents the absolute value, in order to be able to make a decision, with probability at least $1-\delta_{S_{m}}$.

This forms the underlying principle for the creation of a stopping rule \cite{Bardenet2014,Mnih2008}. Let $\delta_s \in (0,1)$ be a user specified input parameter.
The idea is to sequentially increase the size of ${S_{m}}$ while at the same time checking if the stopping criterion, $|\Lambda^{S_{m}}_1(\vc{x}^{m-1},\vc{x}^*)-\psi(\vc{x}^*,\vc{x}^{m-1})|>c_{S_{m}}$, is met. If the stopping criterion is never met, then this will result in ${S_{m}} = {M}$, i.e requiring the evaluation of all the measurements.  Selecting $\delta_{S_{m}} = \frac{p_s-1}{p_s{S_{m}}^{p_s}}\delta_s$ results in $\sum_{{S_{m}}\ge1} \delta_{S_{m}} \leq \delta_s$. The event
\begin{equation}
\mathcal{E} =\!\! \bigcap_{{S_{m}}\ge1}\!\!\left\{\!|\Lambda^{S_{m}}(\vc{x}^{m-1},\vc{x}^*)-\Lambda^{M}(\vc{x}^{m-1},\vc{x}^*)| \leq c_{S_{m}}\!\right\}
\end{equation}
thus holds with probability at least $1-\delta_s$ by a union bound argument.

This iterative procedure allows for an adaptive size of the number of measurements required to be evaluated. However,  there is cause for concern with the definition of the stopping rule. That is the fact that the range, $R$, used in the calculation of \eqref{ct}, is dependent on the log likelihood for all $M$ measurements. Calculating this range would thus inherently require at least the same number of calculations as in the standard sequential MCMC approach. In certain applications it may be possible to obtain an expression for the range which is independent of the measurements, however, this is not the general case. In order to overcome the computational complexity of the calculation of the range, and to reduce the sample variance $V_{S_{m}}$ in the bound, a control variate has been introduced in \cite{Bardenet2015a}, referred to as a proxy:
\begin{equation}
\wp_i(\vc{x}^{m-1},\vc{x}^*) \approx \log\left[\frac{p(\vc{z}^{i}|\vc{x}^*)}{p(\vc{z}^{i}|\vc{x}^{m-1})}\right].
\end{equation}
Thus the MC approximation in \eqref{MCappr} is augmented into
\begin{eqnarray}
\!\!\!\!\!\!\!\!\!\Lambda^{S_{m}}_1(\vc{x}^{m-1},\vc{x}^*)\! =\! \frac{1}{S_{m}}\sum_{i=1}^{S_{m}}\nonumber \log\!\left[\frac{p(\vc{z}^{i,*}|\vc{x}^*)}{p(\vc{z}^{i,*}|\vc{x}^{m-1})}\right]\!\\-\wp_i(\vc{x}^{m-1},\vc{x}^*).\!\!\!\!\!\!\!\!\!
\end{eqnarray}
It is required to amend the MH acceptance accordingly to take the inclusion of the proxy into account.

In \cite{Bardenet2015}, it was proposed to utilise a Taylor series as an approximation for the log likelihood, $\ell_i(\vc{x}) = \log p(\vc{z}^i|\vc{x})$. In this paper we specifically utilise a first order Taylor series,
\begin{equation}
\hat{\ell}_i(\vc{x}) = \ell_i(\vc{x}^+) + (\nabla\ell_i)_{\vc{x}^+}^T\cdot(\vc{x}-\vc{x}^+),
\end{equation}
where $ (\nabla\ell_i)_{\vc{x}^+}$ represents the gradient of ${\ell}_i(\vc{x})$ evaluated at $\vc{x}^+$. This results in the following form of the proxy
\begin{eqnarray}
\nonumber \wp_i(\vc{x}^{m-1},\vc{x}^*)\!\! \!\! \! &= \hat{\ell}_i(\vc{x}^*)-\hat{\ell}_i(\vc{x}^{m-1}), \\ \label{proxycalc}
&=  (\nabla\ell_i)_{\vc{x}^+}^T\cdot(\vc{x}^* - \vc{x}^{m-1}).\!\! \!\! \!\! \!\! \!\! \!\!\!\! \!
\end{eqnarray}
With the inclusion of the proxy, the range, $R$, is now computed as,
\begin{eqnarray}
\nonumber R = \max_{1\le i \le {M}}\left\{\log\left[ \frac{p(\vc{z}^{i}|\vc{x}^*)}{p(\vc{z}^{i}|\vc{x}^{m-1})} \right]-\wp_i(\vc{x}^{m-1},\vc{x}^*)\right\}\\
-\min_{1\le i \le {M}}\left\{\log\left[\frac{p(\vc{z}^{i}|\vc{x}^*)}{p(\vc{z}^{i}|\vc{x}^{m-1})}\right]-\wp_i(\vc{x}^{m-1},\vc{x}^*)\right\}.\!\! \!\!\!\!\! \!\! \!
\end{eqnarray}
We can derive an upper bound for the range, $R^B$, i.e where $R^B~\ge~R$, which can be computed efficiently
\begin{align}
R^B &= 2\max_{1\le i \le {M}}\!\left\{\left|\log\left[ \frac{p(\vc{z}^{i}|\vc{x}^*)}{p(\vc{z}^{i}|\vc{x}^{m-1})}\right]-\wp_i(\vc{x}^{m-1},\vc{x}^*)\right|\right\}\nonumber \\
&= 2\max_{1\le i \le {M}}\!\left\{\left|\ell_i(\vc{x}^*)\! -\!\ell_i(\vc{x}^{m-1}) \!-\!\hat{\ell}_i(\vc{x}^*)\!+\!\hat{\ell}_i(\vc{x}^{m-1})  \right|\right\} \nonumber\\
&= 2\max_{1\le i \le {M}}\!\left\{\left| B({\vc{x}^{m-1}}) - B({\vc{x}^{*}}) \right|\right\}\label{UB_pre}
\end{align}
where $B(\vc{x}) = {\ell}_i(\vc{x}) - \hat{\ell}_i(\vc{x})$ is the remainder of the Taylor approximation. The Taylor-Lagrange inequality  gives us an upper bound on the remainder term. More specifically, if $|f^{(n+1)}(\vc{x})| \le Y$, then $|B_k(\vc{x})| \le \frac{Y|\vc{x}-\vc{x}^+|^{n+1}}{(n+1)!}$, where in our case $n+1 = 2$. Upper bounding the Taylor remainder finally results in the following upper bound on the range
\begin{align}
R^B = 2\left|\left|B({\vc{x}^{m-1}})\right| + \left|B({\vc{x}^{*}})\right| \right|,\label{UB}
\end{align}
which is dependent on the maximum of the Hessian of the log likelihood, $Y$.
The complete adaptive subsampling sequential MCMC approach is illustrated by Algorithms~\ref{SS_SMCMC} and \ref{Confidence_Sampler}.
\begin{algorithm}[!ht]
\caption{Adaptive Subsampling Sequential Markov Chain Monte Carlo}
\label{SS_SMCMC}
\begin{algorithmic}[1]
\small
\STATE Initialize particle set: $\{\vc{x}_0^{(j)}\}_{j=1}^{N}$
\STATE Determine initial proxy parameters. \label{proxparaminit}
\FOR{$k$ = 1,...,$T$}
\FOR{$m$ = 1,...,$N+N_b$}
\IF{$m = 1 \vee N_b$}
\STATE Update proxy parameters. \label{proxparam}
\ENDIF
\STATE \textit{\underline{Joint Draw}}
\STATE Propose $\{\vc{x}_k^*,\vc{x}_{k-1}^*\} \sim q_1\left(\vc{x}_k,\vc{x}_{k-1}|\vc{x}_k^{m-1},\vc{x}_{k-1}^{m-1}\right)$
\STATE Compute $\psi_1(\vc{x}_k^*,\vc{x}_{k-1}^*,\vc{x}_k^{m-1},\vc{x}_{k-1}^{m-1})$ \\ \hspace{2mm}$= \frac{1}{M_k}\log\left[u\frac{p(\vc{x}_k^{m-1}|\vc{x}_{k-1}^{m-1})q_1\left(\vc{x}_k^*,\vc{x}_{k-1}^*|\vc{x}_k^{m-1},\vc{x}_{k-1}^{m-1}\right)}{p(\vc{x}_k^*|\vc{x}_{k-1}^*)q_1\left(\vc{x}_k^{m-1},\vc{x}_{k-1}^{m-1}|\vc{x}_k^*,\vc{x}_{k-1}^*\right)}\right]$
\STATE Compute $\Lambda_1^{S_{m,k}}(\vc{x}_k^*,\vc{x}_k^{m-1})$ and $\{\wp_i(\vc{x}_k^{m-1},\vc{x}_k^*)\}_{i=1}^{M_k}$ with the routine described by Algorithm \ref{Confidence_Sampler}.
\IF{{$\Lambda_1^{S_{m,k}}(\vc{x}_k^*,\vc{x}_k^{m-1}) >\psi_1(\vc{x}_k^*,\vc{x}_{k-1}^*,\vc{x}_k^{m-1},\vc{x}_{k-1}^{m-1})$ \\ \hspace{29mm}$-\frac{1}{M_k}\sum_{i=1}^{M_k}\wp_i(\vc{x}_k^{m-1},\vc{x}_k^*)$}}
\STATE $\{\vc{x}_k^m,\vc{x}_{k-1}^m\} = \{\vc{x}_k^*,\vc{x}_{k-1}^*\}$
\ELSE
\STATE $\{\vc{x}_k^m,\vc{x}_{k-1}^m\} = \{\vc{x}_k^{m-1},\vc{x}_{k-1}^{m-1}\}$
\ENDIF
\STATE \textit{\underline{Refinement}}
\STATE Propose $\{\vc{x}_{k-1}^*\} \sim q_2\left(\vc{x}_{k-1}|\vc{x}_k^{m},\vc{x}_{k-1}^{m}\right)$
\STATE Compute the MH acceptance probability $\rho_2 =$ \\ $\min \left(1,\frac{p(\vc{x}^*_{k-1}|\vc{x}^m_k,\vc{z}_{1:k}) }{q_2\left(\vc{x}^*_{k-1}|\vc{x}_k^{m},\vc{x}_{k-1}^{m}\right)}\frac{q_2\left(\vc{x}_{k-1}^{m}|\vc{x}^m_k,\vc{x}^*_{k-1}\right) }{p(\vc{x}^{m}_{k}|\vc{x}^{m}_{k-1},\vc{z}_{1:k})} \right)$
\STATE Accept $\{\vc{x}_{k-1}^m\} = \{\vc{x}_{k-1}^*\}$ with probability $\rho_2$
\STATE Divide $\vc{x}_k$ into $P$ disjoint blocks $\{\Omega_p\}_{p=1}^P$ such that \\ $\bigcup_p \Omega_p=\{1,...,N_d\}$ and $\Omega_p \cap \Omega_q = \emptyset, \forall p \neq q$
\FOR{$p$ = $1$,...,$P$}
\STATE Propose $\{\vc{x}_k^*(\Omega_p)\} \sim q_{3,p}\left(\vc{x}_k(\Omega_p)|\vc{x}_k^{m},\vc{x}_{k-1}^{m}\right)$
\STATE Compute $\psi_{3,p}(\vc{x}_k^*(\Omega_p),\vc{x}_k^{m}(\Omega_p),\vc{x}_{k-1}^{m})$\\ \hspace{0mm}=$\frac{1}{M_k}\log\left[u\frac{p(\vc{x}_k^{m}(\Omega_p)|\vc{x}_{k-1}^{m})q_{3,p}\left(\vc{x}_k^*(\Omega_p)|\vc{x}_k^{m},\vc{x}_{k-1}^{m}\right)}{p(\vc{x}_k^*(\Omega_p)|\vc{x}_{k-1}^m)q_{3,p}\left(\vc{x}_k^{m}(\Omega_p)|\vc{x}_k^{*},\vc{x}_{k-1}^{m}\right)}\right]$
\STATE Compute $\Lambda_{3,p}^{S_{m,k}}(\vc{x}_k^{m}(\Omega_p),\vc{x}_k^*(\Omega_p))$ and $\{\wp_i(\vc{x}_k^{m}(\Omega_p),\vc{x}_k^*(\Omega_p))\}_{i=1}^{M_k}$ with the routine described by Algorithm \ref{Confidence_Sampler}.
\IF{$\Lambda_{3,p}^{S_{m,k}}(\vc{x}_k^*(\Omega_p),\vc{x}_k^{m}(\Omega_p))$\\  \hspace{10mm}$>\psi_{3,p}(\vc{x}_k^*(\Omega_p),\vc{x}_k^{m}(\Omega_p),\vc{x}_{k-1}^{m})$\\ \hspace{20mm}$-\frac{1}{M_k}\sum_{i=1}^{M_k}\wp_i(\vc{x}_k^{m}(\Omega_p),\vc{x}_k^*(\Omega_p))$}
\STATE $\{\vc{x}_k^m(\Omega_p)\} = \{\vc{x}_k^*(\Omega_p)\}$
\ENDIF
\ENDFOR
\ENDFOR
\STATE Approximation of the marginal posterior distribution with the following empirical measure: \\ $\hat{p}(\vc{x}_k|\vc{z}_{1:k}) = \frac{1}{N}\sum_{j=N_b + 1}^{N+N_b}\delta(\vc{x}_k-\vc{x}_k^{(j)})$
\ENDFOR
\normalsize
\end{algorithmic}
\end{algorithm}

\begin{algorithm}[!ht]
\caption{Adaptive Subsampling Routine}
\label{Confidence_Sampler}
\begin{algorithmic}[1]
\STATE Given: The current and proposed states of the Markov chain, $\{\vc{x}_k$, $\vc{x}_k^*\}$, the complete measurement set, $\vc{z}_k = \{\vc{z}_k^1,...,\vc{z}_k^{M_k}\}$, $\delta$, and $\psi(\cdot)$.
\STATE Initialise: number of sub-sampled measurements, $S_{m,k}~=~0$, Approximate log likelihood ratio subtracted by proxy, $\Lambda~=~0$, set of sub-sampled measurements, $\vc{z}_k^*~=~\emptyset$, initial batchsize, $b~=~1$, while loop counter, $w~=~0$.
\STATE Compute an upper bound for the range, $R_k^B$, according to~\eqref{UB}.
\STATE Compute the proxy, $\{\wp_i(\vc{x}_k,\vc{x}_k^*)\}_{i=1}^{M_k}$, according to \eqref{proxycalc}.
\STATE DONE = FALSE
\WHILE{DONE == FALSE}
\STATE $w = w + 1$
\STATE $\{\vc{z}_k^{S_{m,k}+1,*},...,\vc{z}_k^{b,*}\} \sim_{w/repl.}\vc{z}_k \setminus \vc{z}_k^*$
\STATE $\vc{z}_k^* = \vc{z}_k^* \cup \{\vc{z}_k^{S_{m,k}+1,*},...,\vc{z}_k^{b,*}\}$
\STATE $\Lambda\!\! =\! \!\frac{1}{b}\!\!\left(\!\!S_{m,k}\Lambda\! +\! \!\sum_{i=S_{m,k} + 1}^b\!\! \left[\log \! \frac{p(\vc{z}_{k}^{i,*}\!|\vc{x}_k^*)}{p(\vc{z}_{k}^{i,*}\!|\vc{x}_k)}\!-\!\wp_i(\vc{x}_k,\vc{x}_k^*)\!\right]\!\right)$
\STATE $S_{m,k} = b$
\STATE $\delta_{w} = \frac{p_s-1}{p_s{w}^{p_s}}\delta_s$
\STATE Compute $c$ according to \eqref{ct} utilising $\delta_w$.
\STATE $b = \gamma_s S_{m,k}\wedge M_k$
\IF{$|\Lambda + \frac{1}{M_k}\sum_{i=1}^{M_k}\wp_i(\vc{x}_k,\vc{x}_k^*) - \psi(\cdot)|\ge c$ \OR $S_{m,k} == M_k$}
\STATE DONE = TRUE
\ENDIF
\ENDWHILE
\RETURN $\Lambda$ and $\{\wp_i(\vc{x}_k,\vc{x}_k^*)\}_{i=1}^{M_k}$
\end{algorithmic}
\end{algorithm}
\section{Expectation Propagation Sequential Markov Chain Monte Carlo}
In the previous approach, reduction in computational complexity was based on Bayesian filtering with only a subset of all of the data. In contrast, the algorithm presented in this section utilises all of the data in a distributed way. The only way to achieve computational efficiency is to consider a divide and conquer based approach which processes subsets of the data in parallel. Firstly, the set of $M_k$ measurements is divided into $D$ subsets of measurements such that $\vc{z}_k=\bigcup_{d=1}^D\vc{z}_{k,d}$ and $\vc{z}_{k,i}\bigcap\vc{z}_{k,j} = \emptyset : i \neq j$. The joint filtering  distribution in equation \eqref{eq:jointposterior} is further factored,
{\small
\begin{align}
\Target(\vc{x}_{k},\vc{x}_{k-1}|\vc{z}_{1:k}) \propto p(\vc{x}_k|\vc{x}_{k-1}){\Target}(\vc{x}_{k-1}|\vc{z}_{1:k-1})\prod_{d=1}^{D} p(\vc{z}_{k,d}|\vc{x}_k).\label{fullpost}
\end{align}
}%
The $D$ subsets of measurements are processed in parallel on $D$ computing nodes. The challenge in divide and conquer based approaches is in combining the results from the computing nodes to obtain samples from the joint filtering distribution. A natural method of doing this is through the utilisation of concepts from expectation propagation (EP). EP is a variational message passing scheme \cite{Minka2001}, the EP framework allows for the incorporation of inference from all other $D-1$ computing nodes as a prior in the inference step for any given computing node. This is achieved by approximating the likelihood of the $D-1$ sets of measurements from the other computing nodes with a distribution from the exponential density family,
\begin{equation}
\pi(\state_{k}|\vc{\eta}) = h(\vc{x})g(\vc{\eta})\exp\left\{\vc{\eta}^T\vc{u}(\vc{x})\right\},
\end{equation}
  where $\vc{\eta}$ represents the natural parameters (NPs) and $\vc{u}(\vc{x})$ is a function which varies depending on the member of the exponential family. The local filtering distribution for an individual computing node is thus given by:
\begin{align}
 \Target_{d}(\vc{x}_{k},\vc{x}_{k-1}|\vc{z}_{1:k}) \propto p(&\vc{z}_{k,d}|\vc{x}_k)p(\vc{x}_k|\vc{x}_{k-1})\times\nonumber\\&{\Target}(\vc{x}_{k-1}|\vc{z}_{1:k-1})\prod_{i\neq d}\pi(\state_{k}|\vc{\eta}_i).\label{dpost}
\end{align}
Each local filtering distribution, \eqref{dpost}, is thus an approximation of the joint filtering distribution in \eqref{fullpost}.

The algorithm proceeds iteratively, beginning with the application of MCMC to draw a batch of samples from \eqref{dpost} on each computing node. The NPs of each computing node, $\vc{\eta}_d$, are then determined. This is done by firstly considering the marginalised local filtering distribution,
  \begin{align}
 \Target_{d}(\vc{x}_{k}|\vc{z}_{1:k}) \propto p(\vc{z}_{k,d}|\vc{x}_k)p(\vc{x}_k|\vc{z}_{1:k-1})\prod_{i\neq d}\pi(\state_{k}|\vc{\eta}_i).\label{marg_dpost}
\end{align}
A discrete approximation for the marginalised local filtering distribution can be cheaply obtained from the MCMC samples drawn from the local filtering distribution. Further, by replacing the likelihood expression with the approximate likelihood term, we obtain:
  \begin{align}
 \widehat{\Target}_{d}(\vc{x}_{k}|\vc{z}_{1:k}) \propto \pi(\vc{x}_{k}|\vc{\eta}_d)p(\vc{x}_k|\vc{z}_{1:k-1})\prod_{i\neq d}\pi(\state_{k}|\vc{\eta}_i).\label{marg_dpost_approx}
\end{align}
The idea is to select the NPs, $\vc{\eta}_d$, in a way that results in the minimisation of $\textrm{KL}(\Target_{d}(\vc{x}_{k}|\vc{z}_{1:k})||\widehat{\Target}_{d}(\vc{x}_{k}|\vc{z}_{1:k}))$, where $\textrm{KL}(\, \cdot \,)$ refers to the Kullback-Leibler divergence. It has been shown \cite{Bishop2006} that the minimisation occurs when:
\begin{equation}
\mathbb{E}_{\Target_{d}(\vc{x}_{k}|\vc{z}_{1:k})}\left[\vc{u}(\vc{x})\right]= \mathbb{E}_{\widehat{\Target}_{d}(\vc{x}_{k}|\vc{z}_{1:k})}\left[\vc{u}(\vc{x})\right],
\end{equation}
where $\mathbb{E}\left[\,\cdot\,\right]$ represents the expectation, which corresponds to matching the expected sufficient statistics. Approximating the discrete distributions with the same exponential density family as the likelihood term approximation, i.e.  $\pi(\state_{k}|\vc{\eta}_{p,d}) \approx \Target_{d}(\vc{x}_{k}|\vc{z}_{1:k})$ and $\pi(\state_{k}|\vc{\eta}_{f,d}) \approx {\Target}(\vc{x}_{k}|\vc{z}_{1:k-1}) $, results in the NPs being determined by:
\begin{equation}
\vc{\eta}_d = \vc{\eta}_{p,d} - \left(\vc{\eta}_{f,d} + \sum_{i\neq d} \vc{\eta}_i \right).\label{final_NP_update}
\end{equation}
Finally, the NPs are distributed to all $D \setminus d$ computing nodes, followed by the next iteration.
The number of iterations is dependent on the rate of convergence of $\vc{\eta}_d$ and is treated as a fixed parameter. The EP Sequential MCMC algorithm is described by Algorithm \ref{EPSMCMC}.
\afterpage{%
\begin{algorithm}
\caption{Expectation Propagation Sequential Markov chain Monte Carlo} \label{EPSMCMC}
 \begin{algorithmic}[1]
 \STATE Partition $M_k$ measurements into $D$ sets, $\{\vc{z}_{k,d}\}_{d=1}^D$, and distribute the sets to each corresponding computing node.
 \STATE Initialize particle set on each computing node: $\{\vc{x}_0^{(j)}\}_{j=1}^{N}$
    \FOR{$k=1,...,T$}
        \FOR[EP iteration index]{$L=1,...,L$}
            \FOR[Computing node index (completed in parallel)]{$d=1,...,D$}
            \STATE Follow steps 3 to 19 of Algorithm 1 with equation \eqref{dpost} as the target  distribution.
                \STATE Determine the  NPs of the approximated likelihood term, $\vc{\eta}_d$, according to equation \eqref{final_NP_update}.
                            \STATE Distribute the NPs of the approximated likelihood term to the set $D \setminus d$ computing nodes.
                \ENDFOR
        \ENDFOR
        \STATE Filtering distribution approximated with samples from the $D$ computing nodes.
    \ENDFOR
\end{algorithmic}
\end{algorithm}
}
\section{Proposal Distributions}
The framework presented in this paper for sequential MCMC consists of two sampling stages, referred to as the joint draw and refinement step. However, the framework is flexible in the sense that both sampling stages sample from the target distribution and are thus not both necessarily required for operation. The joint draw has the advantage of only requiring a single evaluation of the measurements. The refinement step introduces additional computational complexity but has also shown to significantly increase the efficiency of the sampling in higher dimensional state space models. Once an appropriate architecture for the sequential MCMC is selected, there is additional flexibility which arises in the form of selection of the proposal distributions. A common choice for the joint draw is to utilise the following proposal distribution:
\begin{equation}
\small q_1\left(\vc{x}_k,\vc{x}_{k-1}|\vc{x}_k^{m-1},\vc{x}_{k-1}^{m-1}\right) = p(\vc{x}_k|\vc{x}_{k-1})\frac{1}{N}\sum_{j=N_b + 1}^{N+N_b}\delta(\vc{x}_{k-1}-\vc{x}_{k-1}^{(j)}).
\end{equation}
In this case, the MH acceptance probability simplifies to a ratio of two likelihoods. This is typically followed by the following proposal distributions for the refinement step:
{\small
\begin{align}
q_2&\left(\vc{x}_{k-1}|\vc{x}_k^{m},\vc{x}_{k-1}^{m}\right) = p(\vc{x}_{k-1}|\vc{x}_k,\vc{z}_{1:k}) \nonumber\\
 &= \sum_{j=\Burnin+1}^{\Burnin+\NbPart} \dfrac{p(\state_{k}=\stateR_{k}^{m}|\stateR_{k-1}^{j})}{ \sum_{i}^{\NbPart} p(\state_{k}=\stateR_{k}^{m}|\stateR_{k-1}^{i})} \delta(\vc{x}_{k-1}-\vc{x}_{k-1}^{(j)}), \label{RefinementPreviousOptimal}
\end{align}
}%
and
\begin{equation}
\small q_{3,p}\left(\vc{x}_k(\Omega_p)|\vc{x}_k^{m},\vc{x}_{k-1}^{m}\right) = p(\vc{x}_k(\Omega_p)|\vc{z}_{1:k},\vc{x}_{k-1},\vc{x}_k(\{1,...,N_d\} \setminus \Omega_p),\label{RefinementCurrentOptimal}
\end{equation}
thus the acceptance ratios $\rho_2$ and $\{\rho_{3,p}\}_{p=1}^P$ will be equal to 1, leading to a refinement stage equivalent to a series of ``perfect'' Gibbs samplers \cite{RobertCasella2004}.

In our case, sampling from equation (\ref{RefinementPreviousOptimal}) is possible at the expense of a large computational cost. Nevertheless the advantage is that this quantity does not depend on the data which is the main challenge in a setting consisting of massive amounts of data. It is also possible to avoid this complexity by using a uniform draw from an index, the acceptance ratio will then reduce to the ratio of two prior distributions.

Typically, sampling from equation (\ref{RefinementCurrentOptimal}) is not possible. Alternatively, the proposal distribution in equation  (\ref{RefinementCurrentOptimal}) can be replaced with a conditional prior or random-walk \cite{Septier2009}. An additional advantage of the EP-SMCMC framework is that each computing node $d$ can utilise the information from the measurements at the other $D \setminus d$ computing nodes in the proposal distribution. It has recently been shown in \cite{Septier2015} how information about the measurements can be utilised in the generic sequential MCMC framework, however, this typically requires additional computations and the evaluation of gradients of the likelihood.

\section{Experiments}
\label{results}
In this section we compare the generic sequential MCMC algorithm with the proposed adaptive subsampling sequential MCMC algorithm, and expectation propagation sequential MCMC algorithm, referred to as SMCMC, AS-SMCMC and EP-SMCMC, respectively. All the algorithms were implemented in the interpreted language Matlab. The parallel processing for the EP-SMCMC algorithm was acheived in Matlab with the \texttt{parfor} command. All simulations were performed on a mobile computer with Intel(R) Core(TM) i7-4702HQ CPU @ 2.20GHz with 16GB of RAM. All results are averaged over 50 independent runs.
\subsection{EP-SMCMC considerations}
For the examples presented in this paper, the member of the exponential family selected to approximate the likelihood terms is the multivariate Gaussian distribution. For this case the NPs are given by:
\begin{equation}
\vc{\eta} = \left(\vc{\Sigma}^{-1}\vc{\mu},\vc{\Sigma}^{-1}\right)^T,
\end{equation}
where $\vc{\mu}$ and $\vc{\Sigma}$ represent the mean and covariance of the multivariate Gaussian distribution. In this case, the NPs update in \eqref{final_NP_update} simplifies to:
 \begin{align}
 \vc{\Sigma}^{-1}_d\vc{\mu}_d &= \vc{\Sigma}^{-1}_{p,d}\vc{\mu}_{p,d} - \left(\vc{\Sigma}^{-1}_{f,d}\vc{\mu}_{f,d} + \sum_{i\neq d} \vc{\Sigma}^{-1}_{i}\vc{\mu}_{i} \right)\nonumber\\
  \vc{\Sigma}^{-1}_d &= \vc{\Sigma}^{-1}_{p,d} - \left(\vc{\Sigma}^{-1}_{f,d} + \sum_{i\neq d} \vc{\Sigma}^{-1}_{i} \right),\label{NPupdate}
 \end{align}
where standard techniques are used to obtain unbiased mean and covariance estimates for the discrete distributions. It is important to note that the difference between two positive definite matrices is not necessarily itself positive definite. Techniques, such as $SoftAbs$ \cite{Betancourt2013}, can be used to ensure that the result remains positive definite.

\subsection{Example 1: Dynamic Gaussian Process with Gaussian likelihood}
The first example is based on a Gaussian state space model with corresponding transition density and likelihood,
\begin{align}
p(\vc{x}_k|\vc{x}_{k-1}) &= \mathcal{N}\left(\state_{k};A\state_{k-1},\vc{Q}\right) \nonumber\\
p(\vc{z}^c_k|\vc{x}_{k}) &= \mathcal{N}\left(\vc{z}^c_{k};H\state_{k},\vc{R}\right).
\end{align}
The measurements are assumed independent, hence resulting in the joint likelihood expression for all measurements:
\begin{align}
p(\vc{z}_k|\vc{x}_{k}) = \prod_{c=1}^{M_k}p(\vc{z}^c_k|\vc{x}_{k}).
\end{align}
The advantage of studying the Gaussian model is that the Kalman filter \cite{Kalman1960} can be used as a benchmark for performance. Unless otherwise specified, the following parameters were utilised for all experiments. The filter parameters include: the number of particles, SMCMC \& AS-SMCMC, $N_p = 4000$, EP-SMCMC, $N_p = 500$ for each computing node (number of computing nodes, $D = 4$); the number of EP iterations, $L = 2$; the subsampling parameters, $\gamma_s = 1.2$, $\delta_s = 0.1$, $p_s = 2$. The simulation parameters include: the number of measurements at each time step, $M = 500$; the total simulation time, $T_{tot} = 20$~s; the transition density parameters, $Q = 0.08$, $A = 0.9$; the likelihood parameters, $H = 1$, $R = 2$; the state space dimension size, $N_d = 1$.

For this example we utilised a sequential MCMC framework consisting of only a refinement step for all three algorithms. In addition, the proposal distribution in \eqref{RefinementPreviousOptimal} was used for the first step in refinement. The conditional posterior for the second refinement step for the SMCMC and AS-SMCMC algorithms is:
\begin{equation}
\small p(\vc{x}_{k}|\vc{x}^m_{k-1},\vc{z}_{1:k}) = p(\vc{z}_k|\vc{x}_k)p(\vc{x}_k|\vc{x}_{k-1}^m).
\end{equation}
The following proposal distribution was selected:
\begin{equation}
\small q_{3}\left(\vc{x}_k|\vc{x}_k^{m},\vc{x}_{k-1}^{m}\right) = p(\vc{x}_k|\vc{x}_{k-1}^m).
 \end{equation}
In the case of EP-SMCMC, the conditional posterior for local computing node $d$ is given by:
\begin{equation}
\small p_d(\vc{x}_{k}|\vc{x}^m_{k-1},\vc{z}_{1:k}) = p(\vc{z}_{k,d}|\vc{x}_k)p(\vc{x}_k|\vc{x}_{k-1}^m)\prod_{i\neq d}\pi(\vc{x}_k|\vc{\eta}_i).
\end{equation}
The following proposal distribution was selected:
 \begin{align}
\small q_{3}\left(\vc{x}_k|\vc{x}_k^{m},\vc{x}_{k-1}^{m}\right) &\propto p(\vc{x}_k|\vc{x}_{k-1}^m)\prod_{i\neq d}\pi(\vc{x}_k|\vc{\eta}_i),\nonumber\\
&= \mathcal{N}\left(\state_{k};\vc{\mu}_q ,\vc{\Sigma}_q\right).
 \end{align}
where $\vc{\mu}_q$ and $\vc{\Sigma}_q$ are derived from the NPs $\vc{\eta}_q = \vc{\eta}_{g,d} + \sum_{i\neq d} \vc{\eta}_i$, and $\vc{\eta}_{g,d}$ represents the NPs of the transition density, $p(\vc{x}_k|\vc{x}_{k-1}^m)$. Table \ref{comp_complex1} illustrates the computational complexity of the algorithms for 500 and 5000 measurements. It is interesting to note that an increase in measurements leads to an increase in computational saving in AS-SMCMC.
\begin{table}
\caption{Algorithm computation time per time step.}
\label{comp_complex1}
\begin{center}
  \begin{tabular}{ | c  |c |  c|}
    \hline
    Algorithms & \multicolumn{2}{|c|}{$M = 500$}
\\ 
\cline{2-3}
 & Time (s)&  \begin{minipage}[t]{0.25\columnwidth}\begin{center}Computational \\ Gain (\%)\end{center}
\end{minipage} \\  \hline
    SMCMC & 114.75 &0 \\  \hline
    AS-SMCMC & 69.54 & 39.4 \\ \hline
    EP-SMCMC  & 9.89 & 91.38\\  \hline
       &\multicolumn{2}{|c|}{$M = 5000$}\\  \hline
          SMCMC & 1087.93  & 0\\  \hline
        AS-SMCMC & 274.60 & 74.76 \\ \hline
       EP-SMCMC  & 96.40 & 91.14\\  \hline
  \end{tabular}
\end{center}
\end{table}

Tables \ref{MCT1} and \ref{MCT2} compare the acceptance rates of the algorithms for the first and second refinement steps, respectively. In Table \ref{MCT1}, the acceptance probabilities for the different algorithms do not differ significantly. This is expected since all three algorithms utilise the same proposal distribution and acceptance ratio for the first refinement step, and additionally, this refinement step is not dependent on the data.
\begin{table}
\caption{Acceptance rates for the first refinement step.}
\label{MCT1}
\begin{center}
  \begin{tabular}{ | c | c |  }
    \hline
    Algorithm& \begin{minipage}[t]{0.4\columnwidth}\begin{center}Acceptance Rate \\ (Min, Median, Mean, Max)\end{center}
\end{minipage}\\ \hline
    SMCMC & (30.93, 94.35, 89.72, 96.57)   \\ \hline
    AS-SMCMC & (30.90, 94.43, 89.70, 96.54)  \\ \hline
    EP-SMCMC (L = 1) & (34.15, 93.99, 89.47, 94.86)  \\ \hline
    EP-SMCMC (L = 2) & (30.86, 94.54, 89.77, 96.59)  \\
    \hline
  \end{tabular}
\end{center}
\end{table}
Table \ref{MCT2} highlights the improvement in acceptance ratio for the EP-SMCMC in this scenario. The increase during the first EP iteration is due to the relative decrease in the number of measurements processed by each computing node. The large increase during the second EP iteration is due to a smarter proposal distribution which incorporates the information about the measurements from the other computing nodes.
\begin{table}
\caption{Acceptance rates for the second refinement step.}
\label{MCT2}
\begin{center}
  \begin{tabular}{ | c | c |  }
    \hline
    Algorithm& \begin{minipage}[t]{0.4\columnwidth}\begin{center}Acceptance Rate \\ (Min, Median, Mean, Max)\end{center}
\end{minipage}\\ \hline
    SMCMC & (8.82, 23.44, 21.26, 25.78)   \\ \hline
    AS-SMCMC & (9.04, 24.24, 21.95, 26.75)  \\ \hline
    EP-SMCMC (L = 1) & (19.76, 42.07, 38.46, 45.01)  \\ \hline
    EP-SMCMC (L = 2) & (72.11, 76.24, 75.86, 77.69)  \\
    \hline
  \end{tabular}
\end{center}
\end{table}

\begin{figure}
\centering
\subfloat[Comparison of the KS statistic for the case of 500 measurements.]{
\includegraphics[width = 80mm]{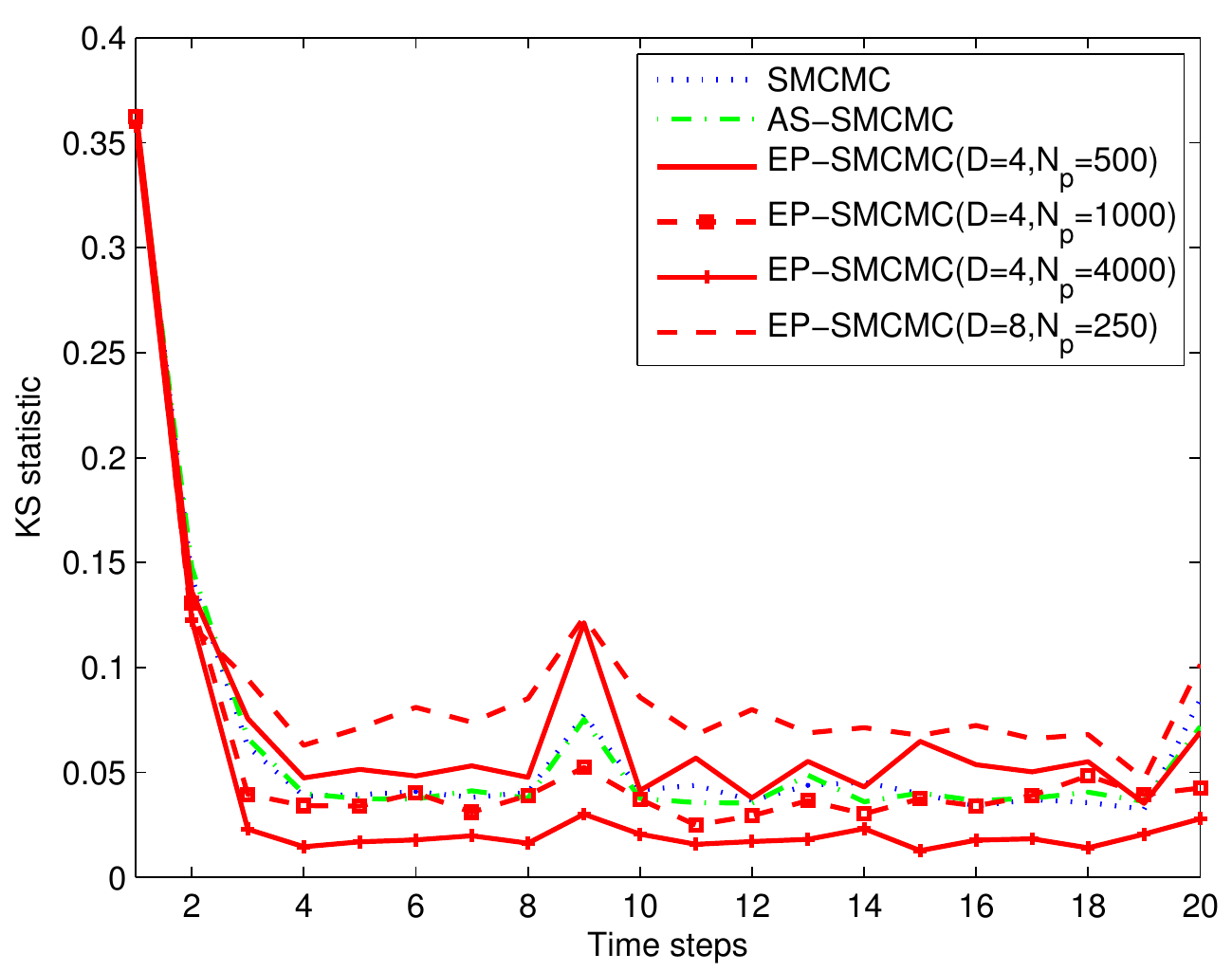}\label{KS500}}
\qquad
\subfloat[Comparison of the KS statistic for the case of 5000 measurements.]{
\includegraphics[width = 80mm]{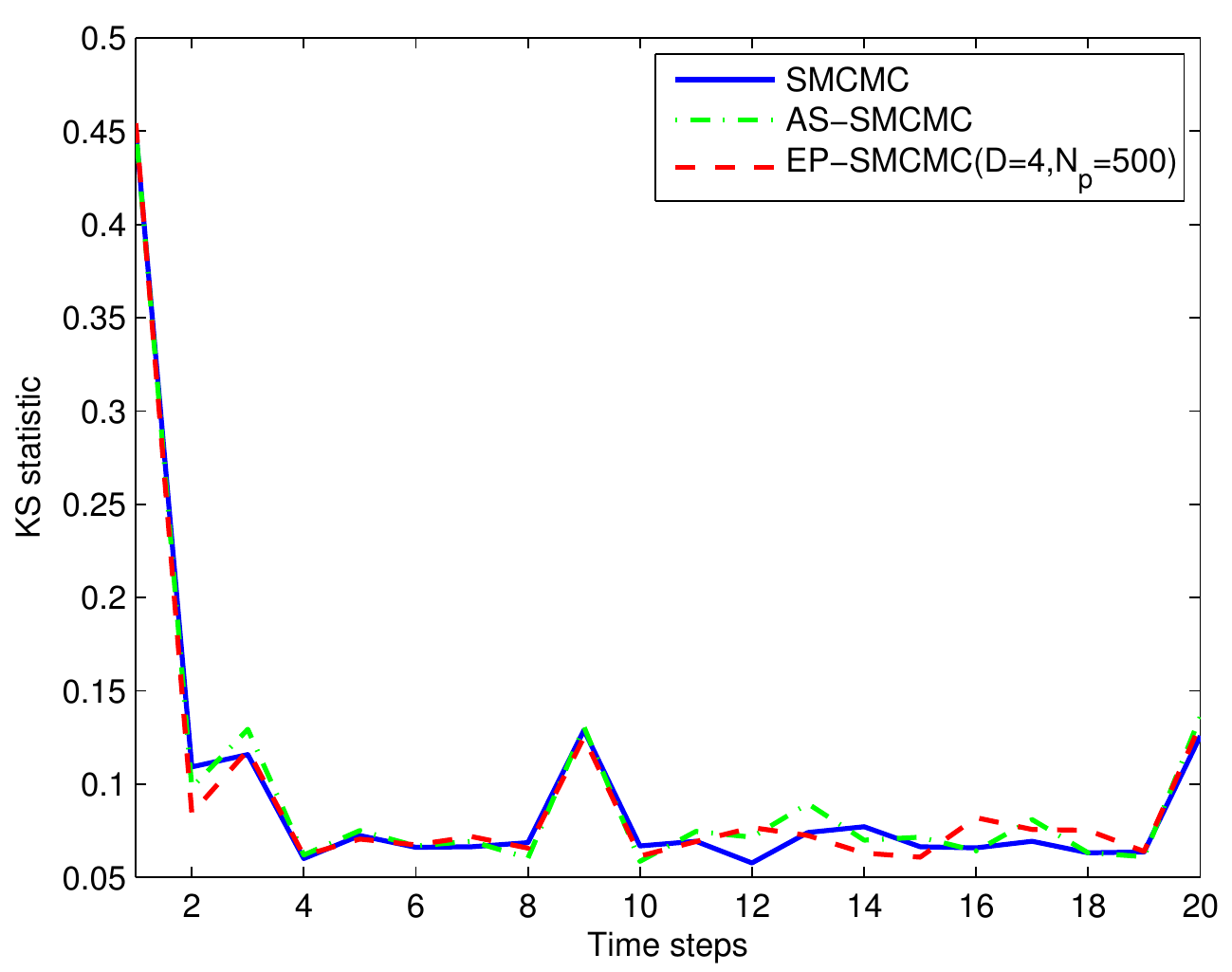}}
\qquad
\caption{The KS statistic for the several configurations of the sequential MCMC based algorithms relative to the Kalman filter.}\label{KScomp}
\end{figure}

 The Kolmogorov-Smirnov (KS) statistic is used to gauge the relative accuracy to correctly approximate empirically the filtering distribution of interest by the algorithms. The KS statistic is given by:
\begin{equation}
KS = \max_{x}\left( \widehat{F}(x) - G(x)\right),
\end{equation}
where $\widehat{F}(x)$ is an empirical cumulative density function (cdf) and $G(x)$ is a continuous cdf. In this setting, $\widehat{F}(x)$ is the empirical cdf of the discrete posterior distribution estimated by the sequential MCMC algorithms, and $G(x)$ the cdf of a Gaussian distribution with parameters updated by a Kalman filter. For EP-SMCMC, the samples from all $D$ computing nodes at the final EP iteration are considered. It is worth while mentioning that the transmission of the samples from the $D$ computing nodes to a single computing node was utilised in this experiment but is not necessary when only estimates are required to be extracted. For example, since the samples in sequential MCMC are unweighted, the global mean can be established through the averaging of the individual local means. The KS statistic for several different filter configurations is illustrated in Figure \ref{KScomp} for both the case of 500 and 5000 measurements. It is first noted that the SMCMC and AS-SMCMC  share almost identical performance. This was expected as the goal of AS-SMCMC is to make the same accept or reject decision in the embedded MCMC algorithms as in SMCMC, only while evaluating less measurements. From Figure \ref{KS500}, we see that the performance of the EP-SMCMC varies depending on the configuration. Doubling the number of computing nodes, while halving the number of samples, conserves the total number of samples while further increasing the computational efficiency at the cost of an increase in error. While in the other extreme case, increasing the number of samples while keeping the number of computing nodes fixed, significantly increases the accuracy while decreasing the computational gain. The case of $N_p$ equal to 1000, results in the same number of samples for all three algorithms. It is clear that even in this scenario, there is an increase in performance, which can be attributed to the increased acceptance rate which results in a more diverse empirical cdf. The EP-SMCMC algorithm is also well suited in this specific example due to the Gaussian nature of the model and utilization of the Gaussian density for the approximate likelihood terms. 

\subsection{Example 2: Multiple Target Tracking in Clutter}
In this example we consider the application of multiple target tracking in clutter. The state vector consists of the positions and velocities of $N_T$ targets in a two dimensional space, $\vc{x}_k~=~[x_{k,1} \hdots x_{k,N_T},y_{k,1}\hdots y_{k,N_T}, \dot{x}_{k,1}\hdots \dot{x}_{k,N_T}, \dot{y}_{k,1}\hdots \\ \dot{y}_{k,N_T}]^T$. In this example it is assumed that the number of targets, $N_T$, is fixed and known, and that each target evolves independently of the other targets. The motion of each target adheres to the near constant velocity model. This results in the marginal state transition density for target $r$ having the form
\begin{equation}
p(\vc{x}_{k,r}|\vc{x}_{k-1,r})=\mathcal{N}(\vc{x}_{k,r}|\vc{A}\vc{x}_{k-1,r},\vc{Q}),
\end{equation}
where $\mathcal{N}(\cdot)$ represents the normal distribution, and matrices $\vc{A}$ and $\vc{Q}$ are defined as $\vc{A} = \left[
                                                           \begin{array}{cc}
                                                             \vc{I}_2 & T_s\vc{I}_2 \\
                                                             \vc{0}_2 & \vc{I}_2 \\
                                                           \end{array}
                                                         \right]$ and $\vc{Q}~=~\sigma^2_x\left[
                                                           \begin{array}{cc}
                                                             (T_s^3/3)\vc{I}_2 & (T_s^2/2)\vc{I}_2 \\
                                                             (T_s^2/2)\vc{I}_2 & T_s\vc{I}_2 \\
                                                           \end{array}
                                                         \right]$, where $T_s = t_k - t_{k-1}$, and $\vc{I}_2$ represents the $2\times2$ identity matrix.

The total number of measurements received is given by $M_k = N_TM_k^x + M_k^c$, where $M_k^x$ represents the number of measurements per target, and $M_k^c$ represents the number of clutter measurements. The number of target and clutter measurements are Poisson distributed with mean $\lambda_X$ and $\lambda_C$ respectively. The likelihood density thus takes the form~\cite{Gilholm2005}:
\begin{equation}
p(\vc{z}_k|\vc{x}_k) = \frac{e^{-\mu_k}}{M_k!}\prod_{i=1}^{M_k} \left( \lambda_Cp_C(\vc{z}_k^i) +\sum_{j=1}^{N_T} \lambda_Xp_X(\vc{z}_k^i|\vc{x}_{k,j}) \right) ,\label{genlik}
\end{equation}
where $\mu_k = \lambda_C + N_T\lambda_T$, $p_X(\cdot)$ and $p_C(\cdot)$ represent the likelihood of a target and clutter measurement respectively. Each individual measurement represents a point in the two dimensional observation space, $\vc{z}_k^i~=~\left[z_{x,k}^i,z_{y,k}^i\right]^T$. In the case of a measurement from a target, the likelihood is modelled as $p_X(\vc{z}_k^i|\vc{x}_{k,j}) = \mathcal{N}(\vc{z}_k^i;\vc{x}_{k,j},\vc{\Sigma})$. The clutter measurements are independent of the states of the targets and are uniformly distributed in the visible region of the sensor, resulting in the clutter likelihood taking the form of $p_C(\vc{z}_k^i)= \textit{U}_{R_x}(z_{x,k}^i) \textit{U}_{R_y}(z_{y,k}^i)$, where $A_c = R_x \times R_y$ represents the clutter area.

The following parameters, unless otherwise specified, were used for all experiments.  The filter parameters include: the number of particles, for SMCMC \& AS-SMCMC, $N_p = 4000$, and EP-SMCMC, $N_p = 500$ for each computing node (number of computing nodes, $D = 4$); the covariance associated with the proposal for the refinement step, $\vc{\Sigma}_r = 0.01\vc{I}$; the subsampling parameters, $\gamma_s = 1.2$, $\delta_s = 0.1$, and $p_s = 2$. The Simulation parameters include: a total running time, $T = 20$, with sampling time, $T_s = 1$; the variance associated with the motion model $\sigma_x = 0.5$; the target observation model parameters, $\lambda_X =1500$, and $\vc{\Sigma} = \vc{I}$; the clutter parameters, $\lambda_C = 4000$, and $A_c = 4\times10^4$; the number of targets $N_T = 3$. 

For this example we utilised a sequential MCMC framework consisting of a joint draw and a local refinement step on the current state only, for all three algorithms. The proposal distribution for the joint draw in the SMCMC and AS-SMCMC is given by:
{\small
\begin{align}
 q_{1}\left(\vc{x}_k,\vc{x}_{k-1}|\vc{x}_k^{m-1},\vc{x}_{k-1}^{m-1}\right) &=  p(\vc{x}_k|\vc{x}_{k-1})\nonumber\\ &\times\frac{1}{N}\sum_{j=N_b + 1}^{N+N_b}\delta(\vc{x}_{k-1}-\vc{x}_{k-1}^{(j)}).
\end{align}}%
The following proposal distribution was selected for the local refinement step:
\begin{equation}
\small q_{3,p}\left(\vc{x}_k(\Omega_p)|\vc{x}_k^{m},\vc{x}_{k-1}^{m}\right) = \mathcal{N}(\vc{x}_k^m(\Omega_p),\vc{\Sigma}_r),\label{local_move}
 \end{equation}
 where $\vc{x}_k(\Omega_p)~=~[x_{k,p}, y_{k,p}, \dot{x}_{k,p}, \dot{y}_{k,p}]^T$ corresponds to the $p$-th target. This proposal represents a random walk move with covariance $\vc{\Sigma}_r$.
In the case of EP-SMCMC, the proposal distribution for the joint draw is given by:
{\small
\begin{align}
q_{1}\left(\vc{x}_k,\vc{x}_{k-1}|\vc{x}_k^{m-1},\vc{x}_{k-1}^{m-1}\right) &\propto  p(\vc{x}_k|\vc{x}_{k-1})\prod_{i\neq d}\pi(\vc{x}_k|\vc{\eta}_i)\nonumber\\ &\times\frac{1}{N}\sum_{j=N_b + 1}^{N+N_b}\delta(\vc{x}_{k-1}-\vc{x}_{k-1}^{(j)})\nonumber\\&= \mathcal{N}\left(\state_{k};\vc{\mu}_q ,\vc{\Sigma}_q\right)\nonumber\\ &\times\frac{1}{N}\sum_{j=N_b + 1}^{N+N_b}\delta(\vc{x}_{k-1}-\vc{x}_{k-1}^{(j)}).
\end{align}}%
where $\vc{\mu}_q$ and $\vc{\Sigma}_q$ are derived from the NPs $\vc{\eta}_q~=~\vc{\eta}_{g,d}~+~\sum_{i\neq d} \vc{\eta}_i$, and $\vc{\eta}_{g,d}$ represents the NPs of the transition density, $p(\vc{x}_k|\vc{x}_{k-1})$.
The same local proposal distribution as used in SMCMC and AS-SMCMC, equation \eqref{local_move}, was selected for the refinement step in EP-SMCMC.

It is interesting to note that in this example the likelihood expression, given in \eqref{genlik}, is independent of a target's velocities. Therefore, when determining the natural parameters of the approximate likelihood terms using \eqref{NPupdate}, the subtraction of the precision terms between the posterior and predictive posterior distributions were forced to zero for all the dimensions related to target velocity. This eliminates potential numerical problems that could arise in the empirical estimation of the natural parameters from a finite number of samples.

\begin{figure}
\centering
\subfloat[True tracks in the $xy$ plane. Start/stop positions are shown with $\bigcirc$/$\Delta$.]{
\includegraphics[width = 65mm]{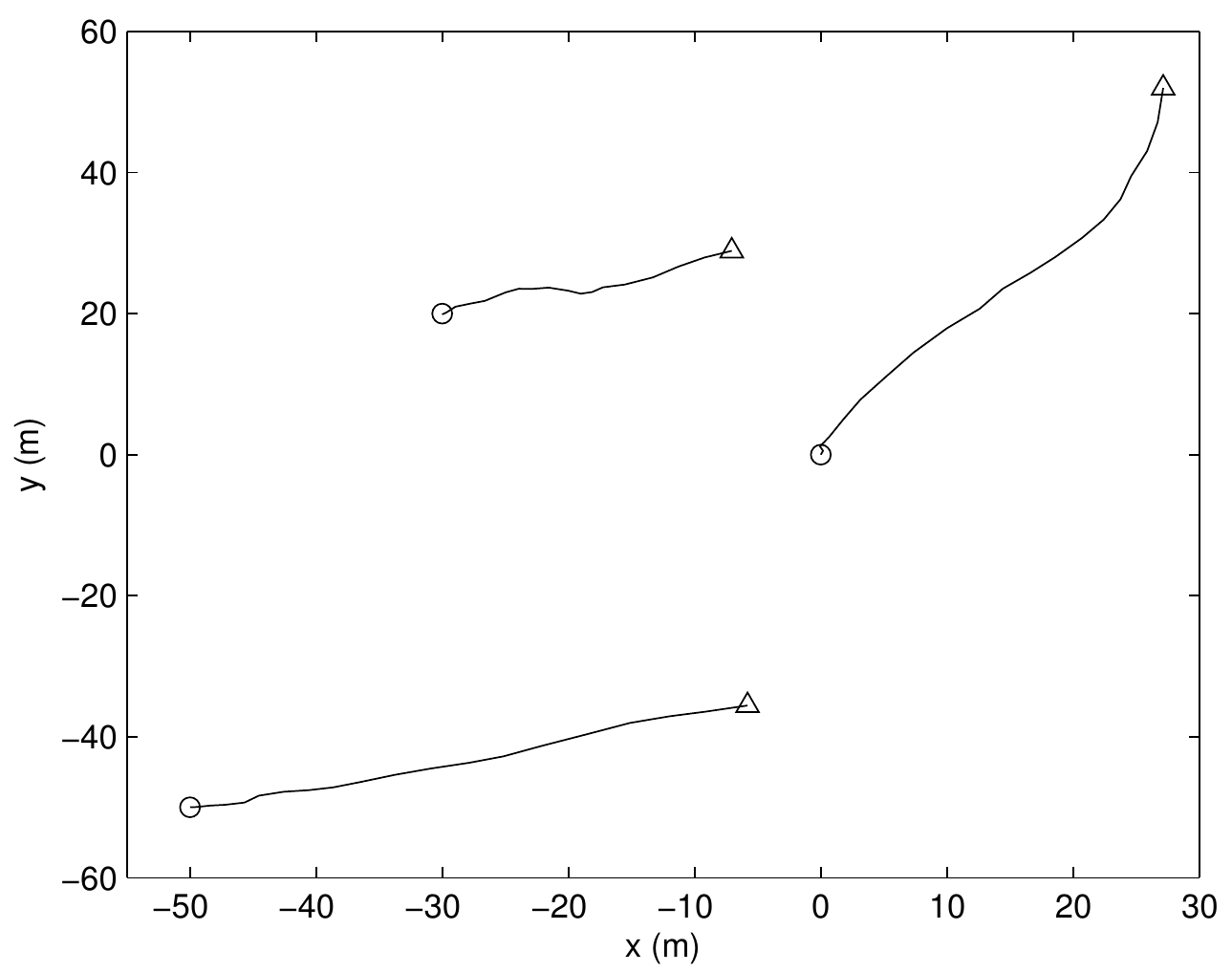}}
\qquad
\subfloat[Tracking result for the generic SMCMC.]{
\includegraphics[width = 65mm]{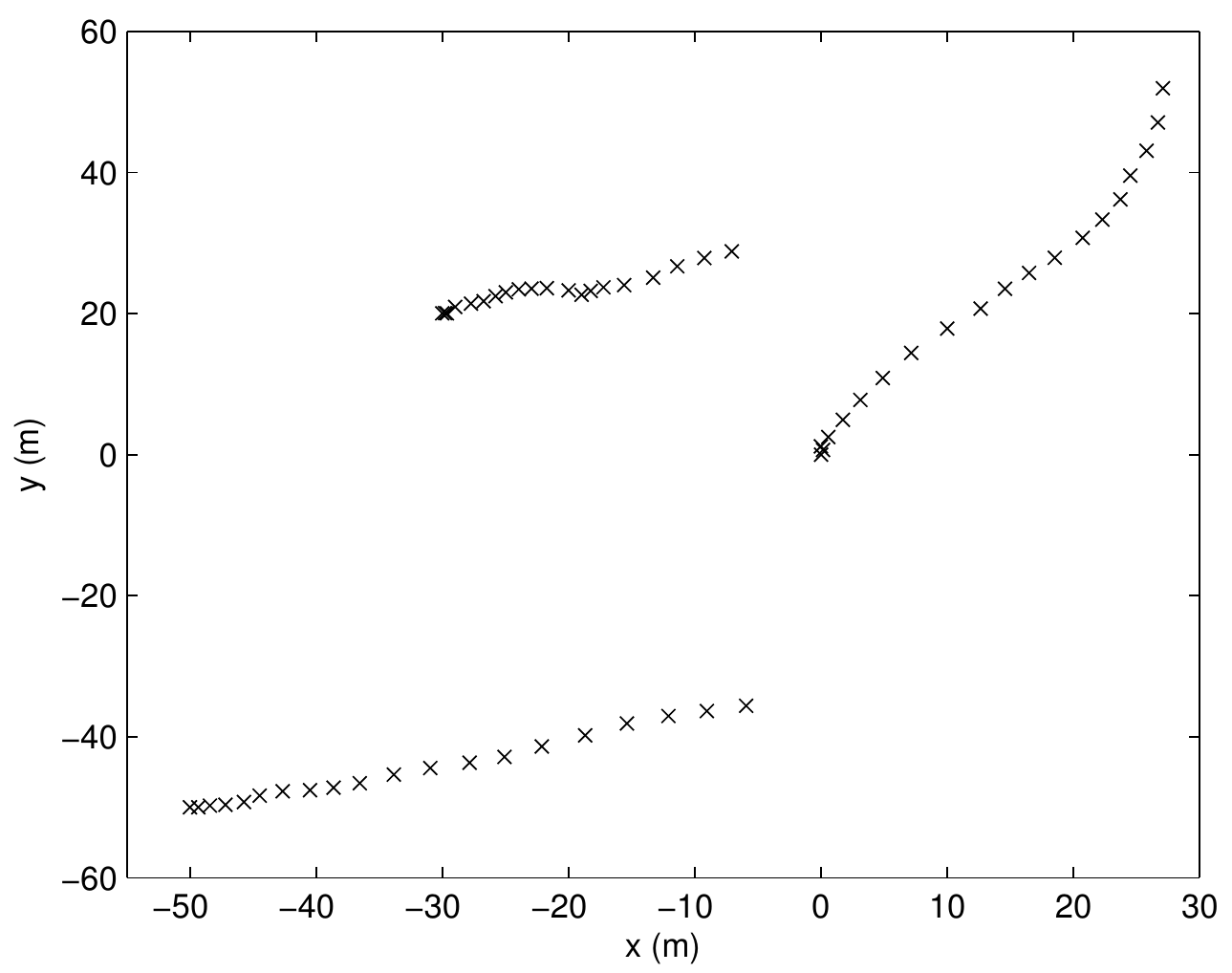}}
\qquad
\subfloat[Tracking result for the AS-SMCMC.]{
\includegraphics[width = 65mm]{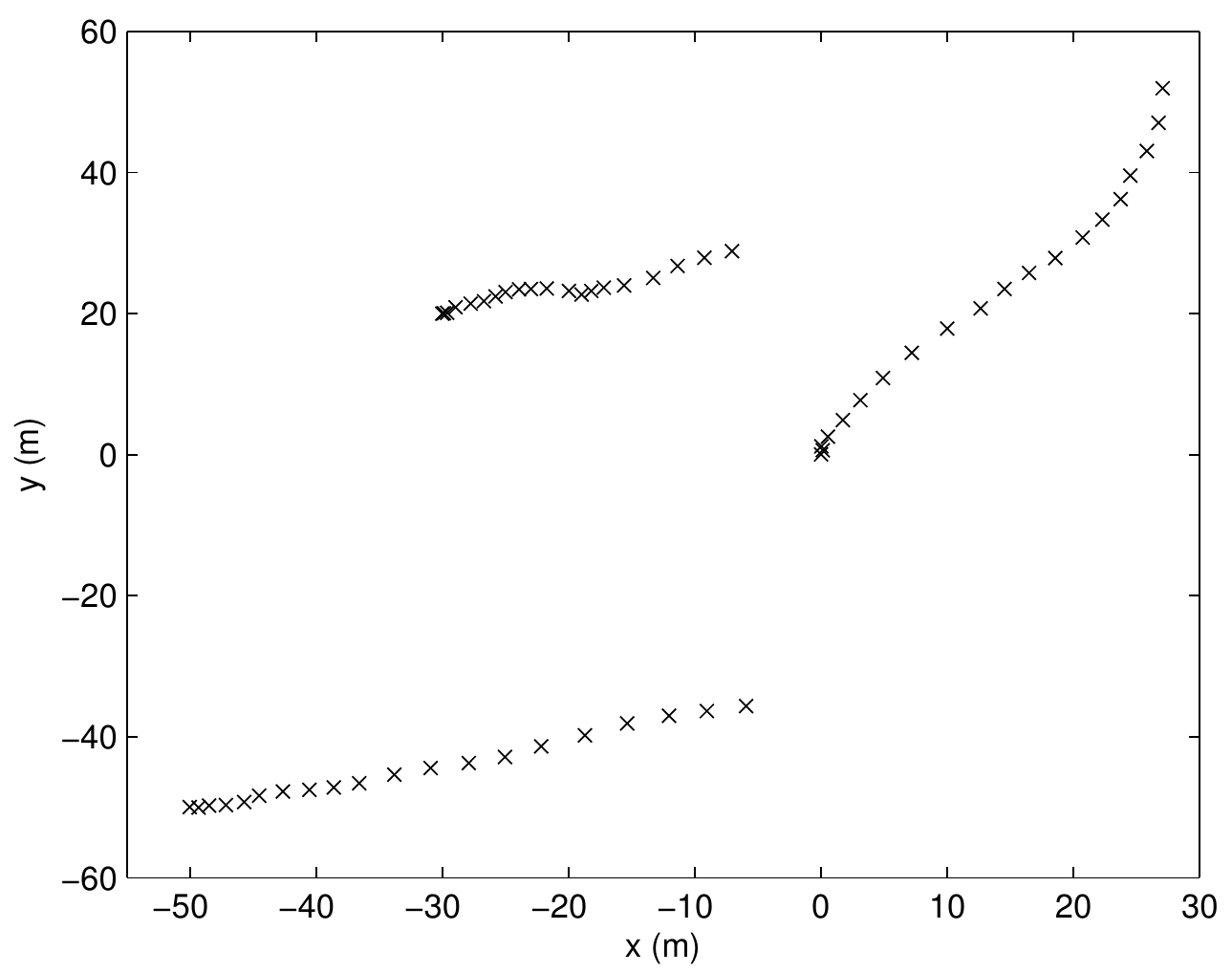}}
\qquad
\subfloat[Tracking result for the EP-SMCMC.]{
\includegraphics[width = 65mm]{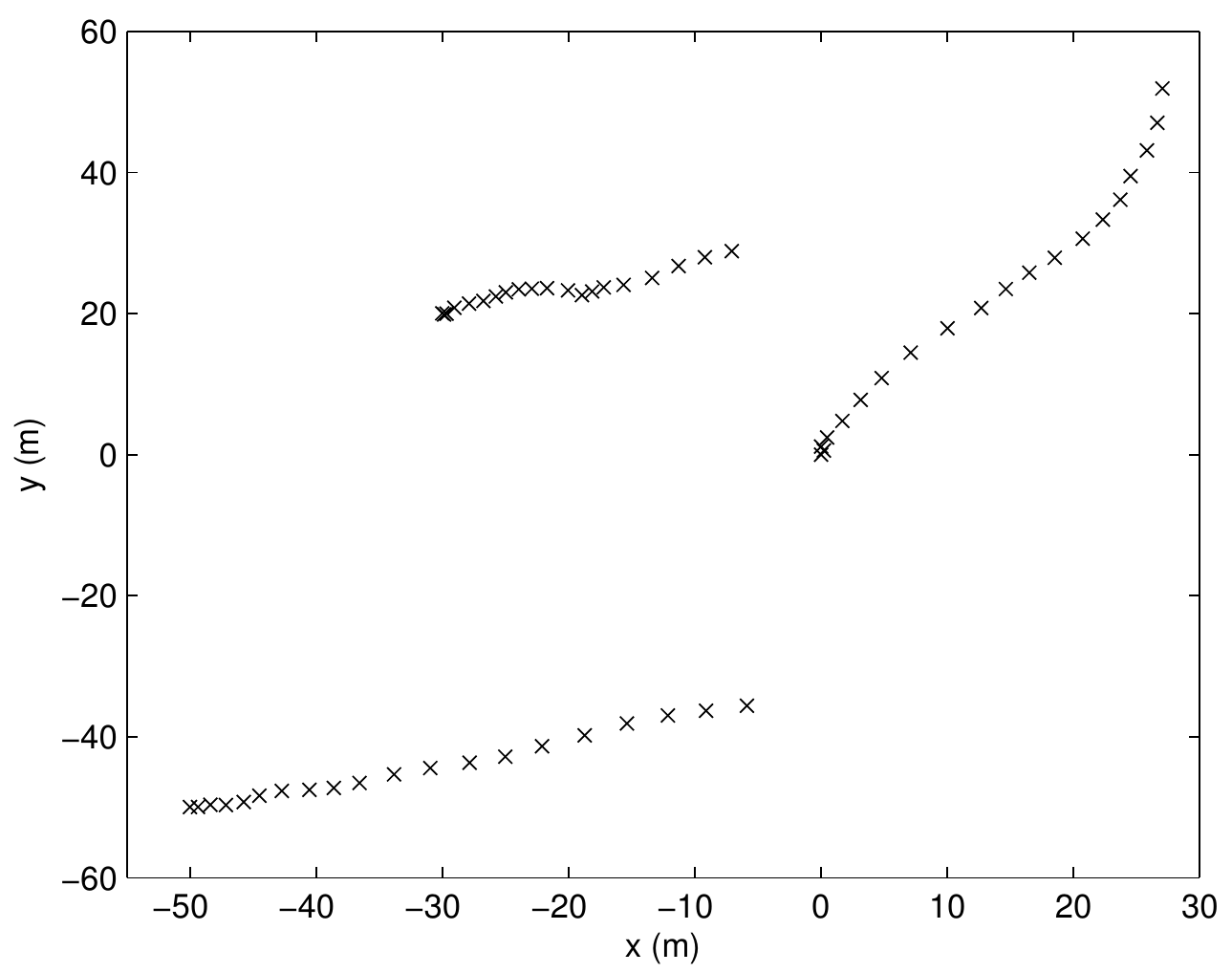}}
\qquad
\caption{Tracking results for a single run of the algorithms.}\label{singlerun}
\end{figure}

\begin{figure}
\includegraphics[width = 80mm]{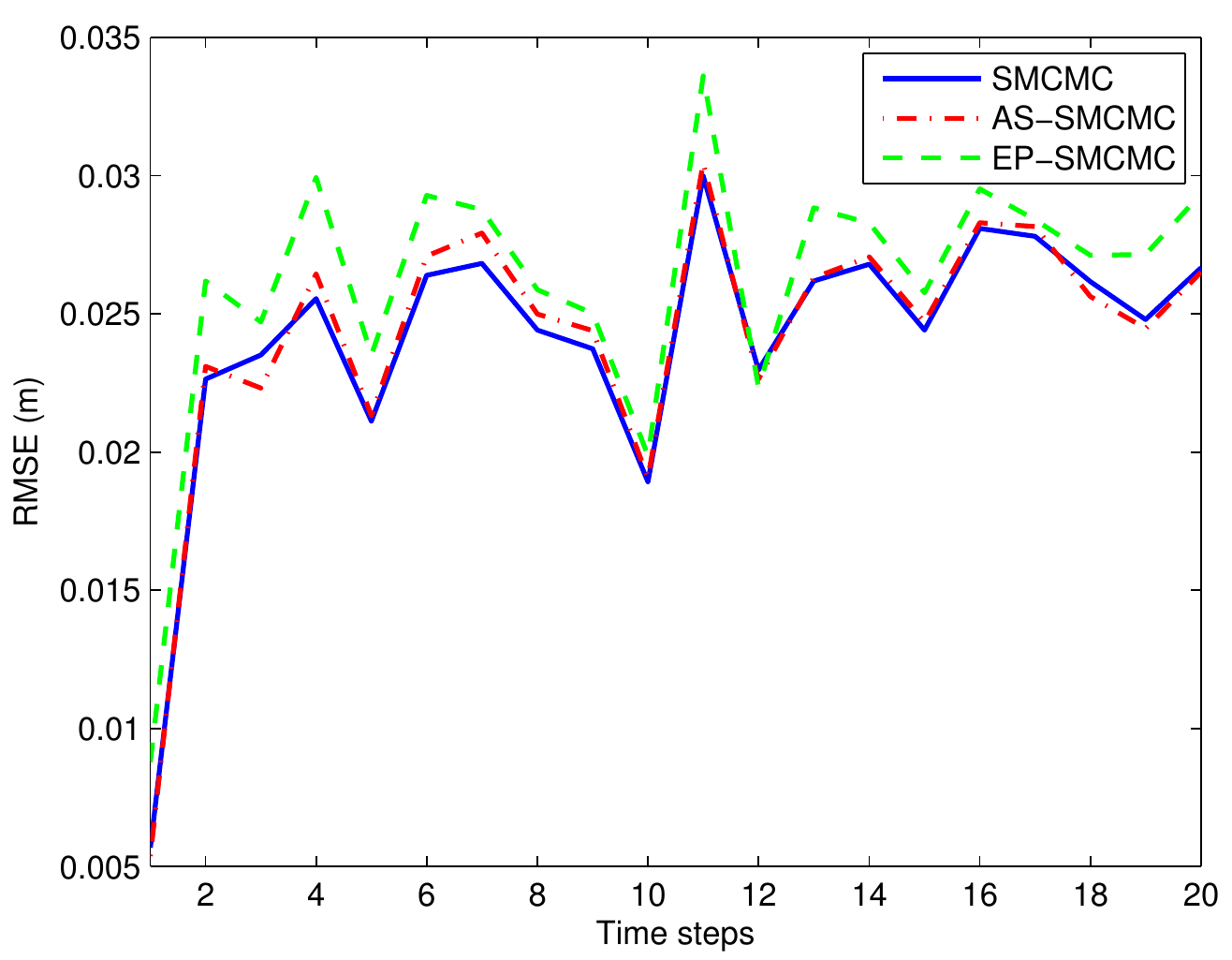}
\caption{The RMSE averaged over the position dimensions for the tracking simulation.}\label{RMSE}
\end{figure}

The estimated tracks for a single simulation run are shown in Figure \ref{singlerun}.
An abundance in measurements results in all algorithms returning accurate estimate results. The root mean square error (RMSE), averaged over all the position dimensions for the three targets, is given in Figure \ref{RMSE}. We found a negligible increase in RMSE for the positions related to the EP-SMCMC. 
The computation time per time step for each algorithm is illustrated in Table \ref{comp_complex2}. The acceptance rates of the joint draw and refinement steps are illustrated in Table \ref{MTT1} and \ref{MTT2} respectively.
\begin{table}
\caption{Algorithm computation time per time step.}
\label{comp_complex2}
\begin{center}
  \begin{tabular}{ | c  |c | c | }
    \hline
    Algorithms & Time (min) & \begin{minipage}[t]{0.25\columnwidth}\begin{center}Computational \\ Gain (\%)\end{center}
\end{minipage}
\\ \hline
    SMCMC & 505.42 &  0\\  \hline
    AS-SMCMC &  388.82 & 23.07  \\ \hline
    EP-SMCMC  & 59.06 & 88.31 \\  \hline
  \end{tabular}
\end{center}
\end{table}
\begin{table}
\caption{Acceptance rates for the joint draw.}
\label{MTT1}
\begin{center}
  \begin{tabular}{ | c | c |  }
    \hline
    Algorithm& \begin{minipage}[t]{0.4\columnwidth}\begin{center}Acceptance Rate \\ (Min, Median, Mean, Max)\end{center}
\end{minipage}\\ \hline
    SMCMC & (0, 0, 0, 0)   \\ \hline
    AS-SMCMC & (0, 0, 0, 0)  \\ \hline 
    EP-SMCMC (L = 1) & (0, 0 , 0.002, 0.01)  \\ \hline
    EP-SMCMC (L = 2) & (0.04, 1.35, 1.35, 3.09)  \\
    \hline
  \end{tabular}
\end{center}
\end{table}
\begin{table}
\caption{Acceptance rates for the refinement step.}
\label{MTT2}
\begin{center}
  \begin{tabular}{ | c | c |  }
    \hline
    Algorithm& \begin{minipage}[t]{0.4\columnwidth}\begin{center}Acceptance Rate \\ (Min, Median, Mean, Max)\end{center}
\end{minipage}\\ \hline
    SMCMC & (28.92, 29.45, 29.45, 30.08)   \\ \hline
    AS-SMCMC & (28.94, 29.31, 29.43, 30.25)  \\ \hline
    EP-SMCMC (L = 1) & (62.37, 63.60, 63.65, 65.43)  \\ \hline
    EP-SMCMC (L = 2) & (25.78, 26.76, 26.92, 28.77)  \\
    \hline
  \end{tabular}
\end{center}
\end{table}
\section{Conclusions}
This paper presents a framework for sequential MCMC. It consists  of  two  sampling  stages,  referred  to  as  the  joint  draw and  refinement  step. The joint draw has the advantage of only requiring
a single evaluation of the measurements. The refinement step
introduces  additional  computational  complexity  but  has  also
shown to significantly increase the efficiency of the sampling
in higher dimensional state space models.We proposed two novel sequential MCMC algorithms capable of greatly reducing the computational time for Bayesian filtering, by up to 90\%. The power of the algorithms was displayed through two examples, with comparisons to a generic sequential MCMC algorithm. The first algorithm is afforded a computational gain by adaptively subsampling the measurements. In contrast, the second algorithm is afforded a computational gain through a divide and conquer approach. Both filters have flexible structures.

\vspace{3mm} \noindent \textbf{Acknowledgments}. We acknowledge the
support from the UK Engineering and Physical Sciences Research
Council (EPSRC) for the support via the Bayesian Tracking and
Reasoning over Time (BTaRoT) grant EP/K021516/1, and [FP7 2013-2017] TRAcking in compleX sensor
systems (TRAX) Grant agreement no.:~607400.

\section*{Appendix}
The AS-SMCMC algorithm requires an upper bound on the range of the log likelihood ratio, as described in equations \eqref{UB_pre} and \eqref{UB}. The upper bound is dependent on the Hessian of the log likelihood. In the examples exhibited in Section \ref{results}, the upper bound is independent of the data and is computed offline prior to tracking. The Hessian for Example 1 is given by:
\begin{equation}
\vc{H} = \vc{R}^{-1},
\end{equation}
and for Example 2:
\begin{equation}
\vc{H} =
\begin{bmatrix}
    \vc{H}_{1,1} & \vc{H}_{1,2} & \vc{H}_{1,3} & \dots  &\vc{H}_{1,N_T} \\
    \vc{H}_{2,1} &\vc{H}_{2,2} & \vc{H}_{2,3} & \dots  & \vc{H}_{2,N_T} \\
    \vdots & \vdots & \vdots & \ddots & \vdots \\
   \vc{H}_{N_T,1} & \vc{H}_{N_T,2} & \vc{H}_{N_T,3} & \dots  & \vc{H}_{N_T,N_T}
\end{bmatrix},
\end{equation}
with 
\begin{align}
\vc{H}_{\ell,\ell} =&\frac{-\lambda_X \vc{\Sigma}^{-1}\mathcal{N}(\vc{z}_k^i;\vc{x}_{k,\ell},\vc{\Sigma})}{\left(\lambda_X \sum_{j}^{N_T}\mathcal{N}(\vc{z}_k^i;\vc{x}_{k,j},\vc{\Sigma}) + \frac{\lambda_C}{A_C}\right)}+\nonumber \\  
&\frac{{\lambda_X} \vc{\Sigma}^{-1}(\vc{z}_k^i-\vc{x}_{k,\ell})\left(\vc{\Sigma}^{-1}(\vc{z}_k^i-\vc{x}_{k,\ell})\right)^T}{\left(\lambda_X \sum_{j}^{N_T}\mathcal{N}(\vc{z}_k^i;\vc{x}_{k,j},\vc{\Sigma}) + \frac{\lambda_C}{A_C}\right)} \times  \nonumber\\
&\frac{\mathcal{N}(\vc{z}_k^i;\vc{x}_{k,\ell},\vc{\Sigma})\left(\lambda_X \sum_{j\neq \ell}\mathcal{N}(\vc{z}_k^i;\vc{x}_{k,j},\vc{\Sigma}) + \frac{\lambda_C}{A_C}\right)}{\left(\lambda_X \sum_{j}^{N_T}\mathcal{N}(\vc{z}_k^i;\vc{x}_{k,j},\vc{\Sigma}) + \frac{\lambda_C}{A_C}\right)},
\end{align}
and 
\begin{align}
\vc{H}_{\ell,j}& =\frac{{\lambda^2_X} \vc{\Sigma}^{-1}(\vc{z}_k^i-\vc{x}_{k,\ell})\left(\vc{\Sigma}^{-1}(\vc{z}_k^i-\vc{x}_{k,j})\right)^T}{\left(\lambda_X \sum_{j}^{N_T}\mathcal{N}(\vc{z}_k^i;\vc{x}_{k,j},\vc{\Sigma}) + \frac{\lambda_C}{A_C}\right)} \times  \nonumber\\
&\frac{\mathcal{N}(\vc{z}_k^i;\vc{x}_{k,\ell},\vc{\Sigma})\mathcal{N}(\vc{z}_k^i;\vc{x}_{k,j},\vc{\Sigma})}{\left(\lambda_X \sum_{j}^{N_T}\mathcal{N}(\vc{z}_k^i;\vc{x}_{k,j},\vc{\Sigma}) + \frac{\lambda_C}{A_C}\right)}.
\end{align}

\bibliographystyle{IEEEtran}
\bibliography{Allan}

\begin{thebibliography}{10}
\providecommand{\url}[1]{#1}
\csname url@rmstyle\endcsname
\providecommand{\newblock}{\relax}
\providecommand{\bibinfo}[2]{#2}
\providecommand\BIBentrySTDinterwordspacing{\spaceskip=0pt\relax}
\providecommand\BIBentryALTinterwordstretchfactor{4}
\providecommand\BIBentryALTinterwordspacing{\spaceskip=\fontdimen2\font plus
\BIBentryALTinterwordstretchfactor\fontdimen3\font minus
  \fontdimen4\font\relax}
\providecommand\BIBforeignlanguage[2]{{%
\expandafter\ifx\csname l@#1\endcsname\relax
\typeout{** WARNING: IEEEtran.bst: No hyphenation pattern has been}%
\typeout{** loaded for the language `#1'. Using the pattern for}%
\typeout{** the default language instead.}%
\else
\language=\csname l@#1\endcsname
\fi
#2}}

\bibitem{Kalman1960}
R.~E. Kalman, ``A new approach to linear filtering and prediction problems,''
  \emph{Trans. of the ASME-Journal of Basic Engineering}, vol.~82, no. Series
  D, pp. 35--45, 1960.

\bibitem{Cappe2007}
O.~Cappe, S.~Godsill, and E.~Moulines, ``{A}n {O}verview of {E}xisting
  {M}ethods and {R}ecent {A}dvances in {S}equential {M}onte {C}arlo,''
  \emph{Proc. IEEE}, vol.~95, no.~5, pp. 899 --924, {M}ay 2007.

\bibitem{Bolic2005}
M.~Bolic, P.~Djuric, and S.~Hong, ``Resampling algorithms and architectures for
  distributed particle filters,'' \emph{IEEE Trans. on Signal Processing},
  vol.~53, no.~7, pp. 2442--2450, July 2005.

\bibitem{Li2015}
T.~Li, M.~Bolic, and P.~Djuric, ``{Resampling Methods for Particle Filtering:
  Classification, implementation, and strategies},'' \emph{IEEE Signal
  Processing Magazine}, vol.~32, no.~3, pp. 70--86, May 2015.

\bibitem{Read2014}
J.~Read, K.~Achutegui, and J.~M\'{i}guez, ``A distributed particle filter for
  nonlinear tracking in wireless sensor networks,'' \emph{Signal Processing},
  vol.~98, no.~0, pp. 121 -- 134, 2014.

\bibitem{Bengtsson2008}
T.~Bengtsson, P.~Bickel, and B.~Li, \emph{{Curse-of-dimensionality revisited:
  Collapse of the particle filter in very large scale systems}}, ser.
  Collections.\hskip 1em plus 0.5em minus 0.4em\relax Beachwood, Ohio, USA:
  Institute of Mathematical Statistics, 2008, vol.~2, pp. 316--334.

\bibitem{VanLeeuwen2014}
P.~{Van Leeuwen}, ``Particle filters for the geosciences,'' in \emph{{Advanced
  data assimilation for Geosciences : lecture notes of the Les Houches School
  of Physics: special issue}}.\hskip 1em plus 0.5em minus 0.4em\relax Oxford:
  Oxford University Press, October 2014, pp. 291--318.

\bibitem{Khan2005}
Z.~Khan, T.~Balch, and F.~Dellaert, ``{MCMC}-based particle filtering for
  tracking a variable number of interacting targets,'' \emph{IEEE Trans. on
  Pattern Analysis and Machine Intelligence}, vol.~27, no.~11, pp. 1805 --1819,
  Nov. 2005.

\bibitem{Septier2009}
F.~Septier, S.~K. Pang, A.~Carmi, and S.~Godsill, ``{On MCMC-Based particle
  methods for Bayesian filtering: Application to multitarget tracking},'' in
  \emph{Proc. of the IEEE Int. Workshop on Computational Advances in
  Multi-Sensor Adaptive Processing}, Dec. 2009, pp. 360--363.

\bibitem{Lamberti2015}
R.~Lamberti, F.~Septier, N.~Salman, and L.~Mihaylova, ``{Sequential Markov
  Chain Monte Carlo for multi-target tracking with correlated RSS
  measurements},'' in \emph{IEEE Tenth International Conference on Intelligent
  Sensors, Sensor Networks and Information Processing}, April 2015, pp. 1--6.

\bibitem{Bardenet2015a}
R.~Bardenet, A.~Doucet, and C.~Holmes, ``{M}arkov chain {M}onte {C}arlo and
  tall data,'' \emph{preprint, http://arxiv.org/abs/1505.02827}, May 2015.

\bibitem{Suchard2010}
M.~Suchard, Q.~Wang, C.~Chan, J.~Frelinger, A.~Cron, and M.~West,
  ``{Understanding GPU programming for statistical computation: Studies in
  massively parallel massive mixtures},'' \emph{Journal of Computational and
  Graphical Statistics}, vol.~19, no.~2, pp. 419--438, 2010.

\bibitem{Scott2013}
S.~L. Scott, A.~W. Blocker, F.~V. Bonassi, H.~A. Chipman, E.~I. George, and
  R.~E. McCulloch, ``{Bayes and big data: The consensus Monte Carlo
  algorithm},'' in \emph{EFaB Bayes 250 Conf.}, vol.~16, 2013.

\bibitem{Neiswanger2013}
W.~Neiswanger, C.~Wang, and E.~Xing, ``{Asymptotically Exact, Embarrassingly
  Parallel MCMC},'' \emph{preprint, http://arxiv.org/abs/1311.4780v2}, 2013.

\bibitem{Casarin2015}
R.~Casarin, R.~Craiu, and F.~Leisen, ``{Embarrassingly Parallel Sequential
  Markov-chain Monte Carlo for Large Sets of Time Series},'' \emph{preprint,
  http://arxiv.org/abs/1512.01496}, 2015.

\bibitem{Wang2013}
X.~Wang and D.~Dunson, ``{Parallel MCMC via Weierstrass sampler},''
  \emph{preprint, http://arxiv.org/abs/1312.4605}, 2014.

\bibitem{Minsker2014}
S.~Minsker, S.~Srivastava, L.~Lin, and D.~Dunson, ``{Robust and Scalable Bayes
  via a Median of Subset Posterior Measures},'' \emph{preprint,
  http://arxiv.org/abs/1403.2660v2}, 2014.

\bibitem{Xu2014}
M.~Xu, Y.~W. Teh, J.~Zhu, and B.~Zhang, ``{Distributed Context-Aware Bayesian
  Posterior Sampling via Expectation Propagation},'' in \emph{Advances in
  Neural Information Processing Systems}, 2014.

\bibitem{Gelman2014}
A.~Gelman, A.~Vehtari, P.~Jylänki, C.~Robert, N.~Chopin, and J.~P. Cunningham,
  ``{Expectation propagation as a way of life},'' \emph{preprint,
  http://arxiv.org/abs/1412.4869}, 2014.

\bibitem{Andrieu2009}
C.~Andrieu and G.~Roberts, ``{The Pseudo-Marginal Approach for Efficient Monte
  Carlo Computations},'' \emph{The Annals of Statistics}, vol.~37, no.~2, pp.
  pp. 697--725, 2009.

\bibitem{Quiroz2014}
M.~Quiroz, M.~Villani, and R.~Kohn, ``{Speeding Up MCMC by Efficient Data
  Subsampling},'' \emph{preprint, http://arxiv.org/abs/1404.4178v1}, 2014.

\bibitem{Bardenet2014}
R.~Bardenet, A.~Doucet, and C.~Holmes, ``{Towards scaling up Markov chain Monte
  Carlo: an adaptive subsampling approach},'' in \emph{Proc. of the Int. Conf.
  on Machine Learning}, 2014, pp. 405--413.

\bibitem{Korattikara2014}
A.~Korattikara, Y.~Chen, and M.~Welling, ``{A}usterity in {MCMC} land:
  {C}utting the {M}etropolis-{H}astings {B}udget,'' in \emph{Proc. of the Int.
  Conf. on Machine Learning}, 2014.

\bibitem{DeFreitas2015}
A.~De~Freitas, F.~Septier, L.~Mihaylova, and S.~Godsill, ``{How can subsampling
  reduce complexity in sequential MCMC methods and deal with big data in target
  tracking?}'' in \emph{Proc. of 18th Int. Conf. on Information Fusion}, July
  2015, pp. 134--141.

\bibitem{Mihaylova2014}
L.~Mihaylova, A.~Carmi, F.~Septier, A.~Gning, S.~Pang, and S.~Godsill,
  ``Overview of {B}ayesian sequential {M}onte {C}arlo methods for group and
  extended object tracking,'' \emph{Digital Signal Processing: A Review
  Journal}, vol.~25, no.~1, pp. 1--16, 2014.

\bibitem{Bardenet2015}
R.~Bardenet and O.-A. Maillard, ``Concentration inequalities for sampling
  without replacement,'' \emph{Bernoulli}, vol.~21, no.~3, pp. 1361--1385,
  2015.

\bibitem{audibert2009}
J.-Y. Audibert, R.~Munos, and C.~Szepesv\'ari, ``Exploration-exploitation
  tradeoff using variance estimates in multi-armed bandits,'' \emph{Theoretical
  Computer Science}, vol. 410, no.~19, pp. 1876 -- 1902, 2009.

\bibitem{Mnih2008}
V.~Mnih, C.~Szepesv\'{a}ri, and J.-Y. Audibert, ``{Empirical Bernstein
  Stopping},'' in \emph{Proc. of the Int. Conf. on Machine Learning}, 2008, pp.
  672--679.

\bibitem{Minka2001}
T.~P. Minka, ``A family of algorithms for approximate {B}ayesian inference,''
  Ph.D. dissertation, Massachusetts Institute of Technology, 2001.

\bibitem{Bishop2006}
C.~M. Bishop, \emph{{Pattern Recognition and Machine Learning}}.\hskip 1em plus
  0.5em minus 0.4em\relax Springer-Verlag New York, 2006.

\bibitem{RobertCasella2004}
C.~P. Robert and G.~Casella, \emph{Monte {C}arlo statistical methods}.\hskip
  1em plus 0.5em minus 0.4em\relax Springer, 2004.

\bibitem{Septier2015}
F.~Septier and G.~Peters, ``{Langevin and {H}amiltonian based {S}equential
  {MCMC} for {E}fficient {B}ayesian {F}iltering in {H}igh-dimensional
  {S}paces},'' \emph{preprint, http://arxiv.org/abs/1504.05715}, 2015.

\bibitem{Betancourt2013}
M.~Betancourt, ``{A general Metric for Riemannian Manifold Hamiltonian Monte
  Carlo},'' \emph{Lecture Notes in Computer Science, Geometric Science of
  Information, Springer}, vol. 8085, no. 327-334, 2013.

\bibitem{Gilholm2005}
K.~Gilholm and D.~Salmond, ``Spatial distribution model for tracking extended
  objects,'' \emph{IEE Proc. Radar, Sonar and Navigation}, vol. 152, no.~5, pp.
  364--371, Oct. 2005.

\end{thebibliography}


%

\end{document}